\documentclass[pdflatex,bst/sn-mathphys-num]{sn-jnl}

\usepackage{amssymb,mathtools,siunitx}
\usepackage{booktabs}
\usepackage{csquotes}
\usepackage{caption}
\usepackage{enumitem}
\usepackage{doi}
\usepackage{indentfirst}

\newcommand{\modifications}[1]{#1}

\def\ab{\boldsymbol{a}}
\def\bb{\boldsymbol{b}}
\def\gb{\boldsymbol{g}}
\def\mb{\boldsymbol{m}}
\def\nb{\boldsymbol{n}}
\def\rb{\boldsymbol{r}}
\def\sb{\boldsymbol{s}}
\def\ub{\boldsymbol{u}}
\def\xb{\boldsymbol{x}}
\def\yb{\boldsymbol{y}}
\def\zb{\boldsymbol{z}}
\def\Ab{\boldsymbol{A}}
\def\Bb{\boldsymbol{B}}
\def\Eb{\boldsymbol{E}}
\def\Fb{\boldsymbol{F}}
\def\Gb{\boldsymbol{G}}
\def\Ib{\boldsymbol{I}}
\def\Mb{\boldsymbol{M}}

\def\Wb{\boldsymbol{W}}
\def\Zb{\boldsymbol{Z}}
\def\Sigmab{\boldsymbol{\Sigma}}
\def\alphab{\boldsymbol{\alpha}}

\def\Pib{\boldsymbol{\Pi}}
\def\D{\mathcal{D}}
\def\E{\mathcal{E}}
\def\F{\mathcal{F}}
\def\G{\mathcal{G}}
\def\H{\mathcal{H}}

\def\N{\mathcal{N}}
\def\P{\mathcal{P}}
\def\X{\mathcal{X}}
\def\opt{\mathrm{opt}}
\def\test{\mathop{\lessgtr}_{\mathcal{H}_1}^{\mathcal{H}_0}}
\def\Pd{P_{\mathrm{d}}}
\def\Pfa{P_{\mathrm{fa}}}
\newcommand{\PdM}[1]{P_{\mathrm{d},#1}}
\newcommand{\PfaM}[1]{P_{\mathrm{fa},#1}}
\def\AUC{\text{AUC}}
\def\pdf{f}
\def\cdf{F}
\def\ccdf{\bar{F}}
\def\FNGS{\Fb_N^{\mathrm{gs}}}
\def\GNGS{\Gb_N^{\mathrm{gs}}}
\newcommand{\sce}[1]{\left( \mathcal{S}_{#1} \right)}
\def\argmax{\operatorname*{argmax}}

\def\Tr{\operatorname{Tr}}
\def\transp{{\top}}
\def\compose{\mathord\circ}
\def\vec{\operatorname{vec}}
\def\vect{\operatorname{span}}

\newcommand{\Esp}[1]{\mathbb{E}\left[ #1 \right]}
\def\SNR{\text{SNR}}
\def\AIC{\text{AIC}}
\def\BIC{\text{BIC}}

\def\Crit{\text{C}}
\def\Maic{M_{\mathrm{aic}}}
\def\Mbic{M_{\mathrm{bic}}}

\def\Mc{M_{\mathrm{c}}}
\def\Rset{\mathbb{R}}
\def\Nset{\mathbb{N}}

\begin{document}

\title{On multipolar magnetic anomaly detection: multipolar signal subspaces, an
  analytical orthonormal basis, multipolar truncation and detection performance}

\author*[1,2,3,4]{\fnm{Cl\'ement}
  \sur{Chenevas-Paule}}\email{clement.chenevas-paule@grenoble-inp.fr}

\author*[1,4]{\fnm{Steeve} \sur{Zozor}}\email{steeve.zozor@cnrs.fr}

\author[2,4]{\fnm{Laure-Line}
  \sur{Rouve}}\email{laure-line.rouve@grenoble-inp.fr}

\author[1,4]{\fnm{Olivier                         J.                         J.}
  \sur{Michel}}\email{olivier.michel@grenoble-inp.fr}

\author[2,4]{\fnm{Olivier} \sur{Pinaud}}\email{olivier.pinaud@grenoble-inp.fr}

\author[3,4]{\fnm{Romain} \sur{Kukla}}\email{romain.kukla@naval-group.com}

\affil*[1]{Univ.  Grenoble Alpes, CNRS, Grenoble INP, GIPSA-Lab, 38000 Grenoble,
  \country{France}}

\affil[2]{Univ.   Grenoble Alpes,  CNRS, Grenoble  INP, G2Elab,  38000 Grenoble,
  \country{France}}

\affil[3]{Centre  d'Expertise  pour  la   Ma\^itrise  de  l'Information  et  des
  Signatures, \orgname{Naval Group}, Ollioules, \country{France}}

\affil[4]{\orgname{Naval Electromagnetism Laboratory}, \orgaddress{21 avenue des
    Martyrs, 38000 Grenoble}, \country{France}}


\abstract{In  this paper,  we consider  the magnetic  anomaly detection  problem
  which aims  to find ferromagnetic  masses by estimating the  weak perturbation
  they induce on local Earth's  magnetic field.  We consider classical detection
  schemes that rely on signals recorded on  a moving sensor, and modeling of the
  source as a  function of unknown parameters.  As the  usual spherical harmonic
  decomposition of  the anomaly has  to be truncated  in practice, we  study the
  signal  vector  subspaces induced  by  each  multipole of  the  decomposition,
  proving they are  not in direct sum,  and discussing the impact it  has on the
  choice of the truncation order. Further,  to ease the detection strategy based
  on generalized likelihood ratio test,  we use orthogonal polynomials theory to
  derive  an analytical  set  of orthonormal  functions (multipolar  orthonormal
  basis   functions)  that   spans  the   space  of   the  noise-free   measured
  signal. Finally, we study the impact of  the truncation order used to tune the
  receiver (generalized likelihood ratio set to a given truncation order) on the
  detection performance,  and the behavior  of the information criteria  used to
  choose this order.}

\keywords{Magnetic    Anomaly   Detection,    Spherical   Harmonic    Multipolar
  Decomposition,  Multipolar  Signal  Subspaces,  Multipolar  Orthonormal  Basis
  Functions, Performance, Truncation and Information Criteria}

\maketitle



\section{Introduction}
\label{Sec:Introduction}


\paragraph{Magnetic anomaly detection in a nutshell}
\label{Subsubsec:MADNutshell}

Magnetic anomaly  detection (MAD\footnote{In  fact, initially, in  the 1940'ies,
the MAD  acronyms referred  to \enquote{Magnetic  airborne detection},  the term
\enquote{airborne}     being     replaced     by     \enquote{anomaly}     later
on~\cite{Fitterman1987MADSurveys}.})   consists in  analyzing  a measured  local
magnetic field to assess  the presence of weak magnetic sources  and is used for
various  applications   (detection  of  underwater  pipes   or  cables,  wrecks,
submarines, etc).  These  sources provoke a local anomaly on  the Earth magnetic
field, that could reveal  their presence~\cite{Zhao2021MADReview}.  MAD is often
based on  the analysis  of a  recorded signal  from a  magnetometer on  board an
aircraft flying  over the ocean  surface, usually following a  linear trajectory
with constant speed  and altitude, while the source is  assumed perfectly still,
at least regarding  the sensor speed (see  Fig.~\ref{Fig:Geometry}).  Usually in
the literature it is assumed that the contribution of the Earth's magnetic field
is subtracted from the measure so that  only the anomaly (corrupted by noise) is
to   be   processed   and  not   the   total   field~\cite{Sheinker2009magnetic,
  Loane1976speed,   Blanpain1979PhD,  Ginzburg2002processing,   Fan2020adaptive,
  Sheinker2007Colorednoise}.   The signal  is  then processed  to determine  the
presence or absence of the source.


\paragraph{A range of paradigms}
\label{Subsubsec:Paradigms}

The  detection of  magnetic anomalies  has  been studied  extensively since  the
second     world      war~\cite{Vacquier1941MADDevice,     Vacquier1941MADNaval,
  MADProgram1946,            Anderson1949MAD,           Fitterman1987MADSurveys,
  Baum1999DetectionObscuredTarget}, right up to the present days, and has led to
the development  of numerous  methods which are  traditionally divided  into $2$
categories~\cite{Zhao2021MADReview}:
\begin{itemize}
  \item source-based approaches,
  \item noise-based approaches.
\end{itemize}
To  these  categories, must  be  added  an  emerging  one, based  on  artificial
intelligence (AI) approaches.

Source-based approaches, to which we will  return at greater length later in the
article, involve establishing  a physical model of the source  thanks to Maxwell
electromagnetism equations and therefore a model of the noise-free signal sensed
on a given trajectory.  The signal described  in this way can be detected using,
for example, Bayesian approaches. Historically, the source has been modeled by a
magnetic              dipole~\cite{Loane1976speed,              Blanpain1979PhD,
  Baum1999DetectionObscuredTarget,    Ginzburg2002processing,   Fan2020adaptive,
  Sheinker2007Colorednoise} whose contribution is predominant at long distances.
The  subtlety provided  by multipole  modeling of  the signal  was only  studied
later~\cite{Pepe2015generalization},  and will  be  explored in  detail in  this
article. \modifications{A  key motivation for  focusing on these methods  is the
  explicit use of  physical constraint in the received signal  model; this makes
  it effective and robust with respect to fluctuations in the noise model, since
  the  representations chosen  on orthogonal  basis functions  --related to  the
  signal model--, induces projections, and algebraic transforms tend to make the
  noise  Gaussian by  virtue  of  the central  limit  theorem.  Furthermore  the
  asymptotic behavior of  the signal model based detector tends  to be robust to
  the noise statistics.}

Noise-based approaches, on the other hand,  make no assumptions about the source
of the anomaly and will focus  on the statistical or entropic characteristics of
the   noise   in   which   an   abrupt    change   is   sought.    By   way   of
illustration,~\cite{Sheinker2008magnetic,  Qiao2023adaptive}  apply  an  entropy
estimation over  a sliding window: if  the entropy falls below  a threshold, the
presence  of  a  source   is  assumed.   \modifications{By  design,  noise-based
  approaches  detect   any  deviation  from  a   reference  noise  distribution,
  regardless  of  the source  of  this  deviation, whether  it  is  due to  some
  unexpected  random perturbation  or the  presence of  a physical  target, thus
  increasing the false alarm rate.}

AI-based approaches  group together  all methods  based on  so-called artificial
intelligence,  which exploit  the  tools  offered by  machine  learning and,  in
particular,     deep    learning     to     carry     out    detection.      For
example,~\cite{Fan2020adaptive}  uses a  support vector  machine (SVM)  which is
applied to  the features extracted from  the signal, namely energy  and entropy.
Similarly,~\cite{Liu2019magnetic} applies  a fully  connected neural  network to
the features  extracted from the  signal.  Many other neural  architectures have
been  used: auto-encoders  for  denoising~\cite{Xu2020deepmad}, residual  neural
networks~\cite{Wang2022deep},  attention   mechanisms,  convolutional  networks,
recurrent   networks~\cite{Wu2021vector,    Hu2020magnetic,   Chen2024magnetic},
etc. Despite the  enthusiasm generated by these techniques, we  need to be aware
that they  are extremely  data-intensive in  order to  carry out  their learning
process, which  is a major constraint.   \modifications{Furthermore, statistical
  performances evaluation and the provision  of detection guarantees have so far
  been extremely challenging, which limits their scope of application.}

\

\modifications{This short introduction  motivates our choice: In  this paper, we
  focus solely on source-based approaches as  the mutual movements of the sensor
  carrier and the  target are presupposed; in addition,  particular attention is
  paid to  the complexity of  the recorded  observation in conjunction  with the
  physical modeling of the source, which will improve detection capabilities.}


\paragraph{From dipolar to multipolar detection}
\label{Subsubsec:DipolarToMultipolar}

Traditionally, source-based  approaches model  the signal  source as  a magnetic
dipole. For  a magnetic  dipolar source at  position $O$ which  will serve  as a
reference, with magnetic  dipolar moment $\Mb$, the magnetic field  at any point
$P$ external to the {\it Brillouin Sphere} (the closest sphere that contains the
source) is given by
\begin{equation*}
\Bb(P) = \dfrac{\mu_0}{4\pi}
\left(
\dfrac{3 \left( \Mb \cdot \rb \right) \rb - r^2 \Mb}{r^5}
\right),  
\end{equation*}
where $\rb$ is  the vector from $O$ to  $P$, $r = \| \rb \|$  its euclidean norm
and  $\cdot$ the  scalar product.   In  a wide  set  of studies,  this model  is
expressed along  the trajectory  so that  the measured signal  lives in  a space
spanned     by      three     basis     functions     known      as     Anderson
function~\cite{Anderson1949MAD,         Loane1976speed,         Blanpain1979PhD}
or~\cite[Chap.~11]{Baum1999DetectionObscuredTarget},  generally  orthonormalized
in what is known as Orthonormal  Basis Functions (OBF).  Algorithms based on OBF
were        widely         studied~\cite{Loane1976speed,        Blanpain1979PhD,
  Ginzburg2002processing, Baum1999DetectionObscuredTarget, Sheinker2009magnetic}
to derive efficient detection methods.  The detection problem usually studied is
that of  a source with  unknown parameters  embedded in additive  white Gaussian
noise, which  leads to  a generalized  likelihood ratio  test (GLRT);  since the
unknowns are in  general the coefficients of the decomposition,  this involves a
projection of the measured signal onto the OBF.  In the far field assumption, it
is often considered that the dipolar approximation is sufficient to describe the
signal effectively.  In a more realistic framework, this assumption is no longer
satisfied  and can  be tackled  with spherical  harmonic (SH)  expansion of  the
magnetic              induction              $\Bb$~\cite{Pepe2015generalization,
  Straton2007Electromagnetic, Jackson1999classical}.


\paragraph{Contribution}
\label{Subsubsec:Contributions}

\modifications{
%
  Section~\ref{Sec:Multipole}  uses   a  direct
approach (avoiding explicit  spherical harmonic expansion) to  recover the basis
fitted  for   finite  order  source   derived  in~\cite{Pepe2015generalization},
spanning  the  space  of  the   multipolar  signal  acquired  along  the  sensor
trajectory.  This allows to identify the signal subspaces generated by each term
of the decomposition;  Their geometrical properties, or the  way they intersect,
shows that, for finite (truncation) order  for the source, the measured magnetic
anomaly may  be represented  by an order  lower than that  of the  source.  This
result is original, up to our knowledge.

Then, a formal description of the binary detection problem for MAD is exposed in
Section~\ref{Sec:MAD}.

We  report  in   Section~\ref{Sec:MOBF}  a  new  analytical   derivation  of  an
orthonormal basis,  based on orthogonal  polynomial theory, which will  be named
MOBF for {\it  multipolar orthonormal basis functions}.   This original approach
provides the  advantage of disposing  of a  close form expression  thus avoiding
possibly  unstable numerical  orthonormalization  processes.  This is  discussed
together with  the impact  of sampling  the signal on  the constructed  MOBF, to
conclude the section.

Finally, in  Section~\ref{Sec:MMAD}, we provide  new insights in  the analytical
study of the performance of the  generalized likelihood ratio test (GLRT) in the
situation where  the measured  anomaly is corrupted  by additive  white Gaussian
noise, especially  in the light of  the question of the  signal truncation order
considered  for  the receiver  setup  and  the  relationships between  the  {\it
  multipole subspaces}.
  
We additionally  experiment and evaluate various  theoretic information criteria
in   order  to   determine  an   adequate  order   to  tune   the  receiver   in
Section~\ref{Sec:Criteria}. Developing information criterion based approaches in
the MAD framework shed a new light on the model complexity selection.

The  very  last  section is  devoted  to  a  conclusion,  allowing to  open  new
discussions and perspective.
}

\section{Multipolar source modeling and multipolar signal subspaces}
\label{Sec:Multipole}


\subsection{Magnetostatic equation}
\label{Subsec:Magnetostatic}

In the present MAD framework, the sensor  is assumed to evolve at positions that
remain outside the Brillouin sphere.  In addition, all sources are assumed to be
perfectly still,  constant in time (on  the scale of the  observation duration).
As  a consequence  of the  Maxwell-Thomson and  Maxwell-Amp\`ere equations,  the
magnetostatic field  $\Bb(P)$ produced by the  sought source at position  $P$ in
source-free  space can  be expressed  using  the scalar  potential $\Psi(P)$  as
follows~\cite{Straton2007Electromagnetic, Jackson1999classical}:
\begin{equation*}
\Bb(P) = -\nabla\Psi(P),
\end{equation*}
where $\Psi(P)$ satisfies the classical Laplace Equation
\begin{equation*}
\Delta \Psi(P) = 0.
\end{equation*}
The resolution  of this equation  is usually based on  a separation-of-variables
method in  spherical coordinates (since  the problem is  spherically symmetric),
that        enables        the        solution       to        be        derived
analytically~\cite{Courant1962tMethodsOfMathematicalPhysicsV2,
  Straton2007Electromagnetic,  Jackson1999classical}.    In  the  rest   of  the
article, we consider both spherical  or Cartesian (in future section) coordinate
systems,  centered on  the  origin  of the  Brillouin  sphere.  Solving  Laplace
equation in spherical coordinates $( r,  \theta , \phi )$, outside the Brillouin
sphere,  leads to  the  expression below,  which exhibits  an  expansion of  the
magnetic            scalar           potential            in           spherical
harmonics~\cite{Straton2007Electromagnetic, Jackson1999classical}:
\begin{equation*}
\displaystyle \Psi(P) = \sum_{l \in \Nset^*} \Psi^{(l)}(P),\\[2mm]
\end{equation*}
with
\begin{equation}
\begin{split}
\Psi^{(l)}(P) =
\dfrac{\mu_0}{4  \pi}   \dfrac{1}{r^{l+1}}  \sum_{m=0}^{l}   \Big(a_{lm}  \cos(m
\varphi) + b_{lm} \sin(m \varphi) \Big) \, \P_l^m(\cos \theta).
\end{split}
\label{Eq:harmo-spher}
\end{equation}
$\Psi^{(l)}(P)$ is called the  {\it magnetic potential of multipolar\footnote{$l
  =1$ corresponds  to the  dipole, $l=2$ to  the quadrupole,\ldots}  order} $l$;
$\P_l^m$ is  the Associated  Legendre polynomial  of degree  $l$ and  order $m$.
Note that, as the solution is  established outside the Brillouin sphere, all the
divergent  source terms  (given for  $l <  0$) have  disappeared from  the above
expression; the monopole term (given for $l=0$) cancels as well, since it has no
physical reality;  the resulting potential  is referred  to as the  {\it partial
  scalar  potential}. Another  expression for  the partial  scalar potential  is
commonly      used~\cite{Pirani1965introduction,      Guth2012TracelessLegendre,
  Guth2012TracelessSpherical,  Gonzalez1998multipole},  which gives  an  insight
into  the  structure  of  the   field.   Not  only  this  facilitates  numerical
simulations,  but also  provides a  simple expression  for the  partial magnetic
field. Firstly, the partial magnetic potential  is reformulated as a function of
a traceless symmetric tensor\footnote{Symmetric means  invariant by any of $l! $
possible permutations $\sigma$  of indices $m_{\sigma(i_1),\ldots,\sigma(i_l)} =
m_{i_1,\ldots,i_l}$; without trace means that if  two indices are equal, the sum
over  these indices  is  zero, i.e.,  the partial  magnetic  potential, or  pure
$2^l-$polar field (dipolar for $l=1$,  quadrupolar for $l=2$,\ldots), is written
as  a function  of  a symmetrical  tensor without  trace,  by symmetry,  $\sum_i
m_{i,i,i_3,\ldots,i_l} = 0$. } as
\begin{equation}
\label{Eq:MagneticMoments}
\Psi^{(l)} = \dfrac{\mu_0}{4 \pi} \dfrac{\rb . \Mb^{(l)}}{r^{l+2}},
\:\:\:  \Mb^{(l)} =  \frac{\mb^{(l)} \,  *_{l-1} \,  \rb^{\otimes (l-1)}}{l!  \,
  r^{l-1}},
\end{equation}
where the  dependence in $P$  is omitted  for simplicity, and  with $\rb =  r \,
\ub_r$ where  $\ub_r$ represents  the unit  radial vector,  and within  a factor
$1/l!$, $\Mb^{(l)}$ is a vector resulting from the Einstein product\footnote{For
a $p-$order tensor $\ab$ and a  $q-$order tensor $\bb$, $n \leqslant \min(p,q)$,
the Einstein product $\ab *_n \bb$ is a $(p+q-2 n)-$order tensor with components
$\sum_{j_1,\ldots,j_n}                     a_{i_1,\ldots,i_{p-n},j_1,\ldots,j_n}
b_{i_{p-n+1},\ldots,i_{p+q-2                               n},j_1,\ldots,j_n}$.}
$*_{l-1}$~\cite{Brazell2013solving,   Lai2010continuummechanics}    between   an
$l-$order symmetric traceless  tensor $\mb^{(l)}$ of size $3$  in each dimension
and  the  unitary  $\ub_r  =  \frac{\rb}{r}$  to  the  $(l-1)-$th  tensor  power
($\otimes$ denotes the tensorial product).  For  a given order $l$, the harmonic
coefficients  $a_{lm}$,   $b_{lm}$  are  one-to-one  mapped   with  the  tensors
$\mb^{(l)}$.   We let  the reader  to Ref.~\cite{Pirani1965introduction}  for an
expression  of the  tensor  based  on the  harmonic  coefficients.  The  partial
magnetic field of order $l$, or pure $2^l-$polar magnetic field, is defined by
\begin{equation}
\label{Eq:partial-field}
\Bb^{(l)}(P)= -\mathbf{\nabla} \Psi^{(l)}(P).
\end{equation}
This calculation was performed in~\cite{Kielanowski1998exact} and gives
\begin{equation}
\label{Eq:MagneticMultipole}
\Bb^{(l)}(P) = \dfrac{\mu_0}{4 \pi} \,  \dfrac{(2l+1) \left( \rb \cdot \Mb^{(l)}
  \right) \rb - l \, r^2 \Mb^{(l)}}{r^{l+4}}.
\end{equation}
The total magnetic field is then
\begin{equation}
\label{Eq:MagneticField}
\Bb(P) = \sum_{l \in \Nset^*} \Bb^{(l)}(P).
\end{equation}


\subsection{Expression of the field along the sensor trajectory}
\label{Subsec:AlongTheTrajectory}


\subsubsection{Geometry, notations}
\label{Subsubsec:Geometry}

\begin{figure}[htbp]
\begin{minipage}{.5\linewidth}
  \centering
  \includegraphics[width=.82\linewidth]{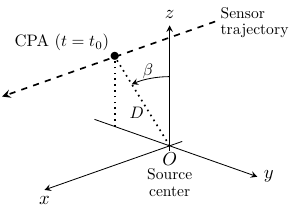}
\end{minipage}
\begin{minipage}{.5\linewidth}
  \caption{Geometry of the  problem. Center of the source is  located at $O$ and
    sensor's position  $P$ moves  along the  dashed line.   $D$ is  the distance
    between the CPA and the source location, and is reached at time $t_0$ by the
    sensor.}
  \label{Fig:Geometry}
\end{minipage}
\end{figure}

In this paper, MAD is assumed to  be performed from recording the magnetic field
on a sensor following a linear trajectory of constant altitude. Furthermore, the
velocity of the sensor on the trajectory  is assumed constant. All points of the
trajectory are outside the Brillouin sphere.  In order to have simple equations,
a Cartesian coordinate system $\{ 0, x,  y, z \}$ centered on the source (center
of the Brillouin sphere) is introduced.   The $z-$axis denotes the vertical axis
so that  the $xy-$plane is the  horizontal plane.  The trajectory  is assumed to
remain parallel  to $x-$axis without loss  of generality.  The closest  point to
the source origin is  called CPA (Closest Point of Approach).   Let $t_0$ be the
instant when  the sensor is at  the CPA, $D$ the  minimal source-sensor distance
and  $\beta$   the  angle  made  by   the  line  ($O-$CPA)  with   the  vertical
axis. Fig.~\ref{Fig:Geometry} summarizes these notations and assumptions.

It is emphasized here that neither $t_0$  nor $D$ (or simply stated the CPA) are
known in practical situations.  In the remainder of this paper, these quantities
will, however, be considered as fixed known parameters.


\subsubsection{Deriving a basis for the field on the trajectory}
\label{Subsubsec:Basis}

In the Cartesian coordinate system introduced  above, $\rb = \begin{bmatrix} x &
  -D\sin\beta  & D\cos\beta  \end{bmatrix}^\transp$, and  $r =  \sqrt{x^2+D^2}$.
From  Eq.~\eqref{Eq:MagneticMoments},  it  is  inferred that  the  numerator  of
$\Mb^{(l)}$  is  a  vector  whose   components  are  degree  (at  most)  $(l-1)$
polynomials  in $x$.   Thus, from  Eq.~\eqref{Eq:MagneticMultipole}, both  terms
from the numerator  of $\Bb^{(l)}(P)$ are vectors polynomials of  degree at most
$(l+1)$  in  this  same  variable,  while the  denominator  is  $r^{2l+3}$.   We
introduce the  unitless reduced variable $u  = \frac{x}{D}$, thus $r  = D \left(
1+u^2 \right)^{\frac12}$,  and the expression  of the pure  $2^l-$polar magnetic
field along the trajectory can be expressed as
 \begin{equation}
\label{Eq:MagneticMultipoleTrajectory}
\Bb^{(l)}(u)    =   \dfrac{{\displaystyle    \sum_{n=0}^{l+1}}   \alphab_n^{(l)}
  u^n}{\left( 1 + u^2 \right)^{l + \frac32}},
\quad u = \dfrac{x}{D} = \dfrac{V \left( t - t_0 \right)}{D},
\end{equation}
where  the 3-dimensional  vector coefficients  $\alphab_n^{(l)}$ depend  only on
tensor $\mb^{(l)}$ and angle $\beta$.

In  order to  derive the  signal  measured by  the sensor  along the  trajectory
(parallel to  $x-$axis), additional  assumptions are  necessary. Firstly,  it is
supposed that the  higher multipolar order is $L$. Either  for physical reasons,
or  because  the  approximation  obtained  by truncating  the  signal  model  at
multipolar order $L$ is satisfactory; remind  that the multipolar field of order
$l$  decreases like  $\frac{1}{r^{l+2}}$.  Secondly, the  sensor  is assumed  to
measure the projection of the vectorial  anomaly along $d$ directions, which are
assumed invariant  during the displacement  (e.g., three-axes sensor,  with axes
not necessarily aligned with the trajectory, but that does not rotate during the
displacement)\footnote{For scalar sensors measuring  a modulus, when the Earth's
field contribution is  removed, since the anomaly is weak,  a good approximation
of  the measurement  is the  projection of  the 3D  anomaly to  the unit  vector
defined by the  Earth's field direction~\cite{Zhao2021MADReview, Loane1976speed,
  Ginzburg2002processing}    or~\cite[\S~11.2]{Baum1999DetectionObscuredTarget},
that  is naturally  constant during  the acquisition.\label{Foot:ScalarSensor}};
this insures  that the projections  of the coefficients $  \alphab_n^{(l)}$ onto
the  sensing directions  remain constant  when $u$  varies.  Let  $\Pib$ be  the
sensing  projection matrix\footnote{For  a  three orthogonal  axes sensor,  this
matrix is just a rotation matrix for  instance; for a scalar sensor as described
in footnote~\ref{Foot:ScalarSensor},  this matrix is  the transpose of  the unit
vector    having     the    Earth's    field    direction.}.      Then,    using
equations~\eqref{Eq:MagneticField}                                           and
Eq.~\eqref{Eq:MagneticMultipoleTrajectory}, and  reducing all terms to  the same
denominator, the signal recorded along the trajectory is expressed as
\begin{equation*}
\sb(u)  =   \dfrac{\displaystyle  \Pib   \,  \sum_{l  =   1}^L  \sum_{n=0}^{l+1}
  \alphab_n^{(l)} u^n (1+u^2)^{L-l }}{(1+u^2)^{L+\frac32}}.
\end{equation*}
The numerator of $\sb(u)$ above is a sum  of polynomials of degree $2 \, (L-l) +
n$ in $u$, thus resulting in a polynomial of maximum degree $2 \, L$ in $u$.  By
inserting  $\Pib$ inside  the sum  and reordering  the terms  in the  expression
above, an alternate expression of $\sb(u)$ is obtained:
\begin{equation}
\label{Eq:Signal}
\sb(u) = \sum_{n=0}^{2 \, L} \ab_L^{(n)} \, f_{L,n}(u),
\end{equation}
where
\begin{equation}
\label{Eq:SignalBasis}
\begin{array}{cccl}
f_{L,n}  : & \Rset & \rightarrow & \Rset\\
& u & \mapsto & \dfrac{u^n}{\left( 1 + u^2 \right)^{L + \frac32}}.
\end{array}
\end{equation}
and $\ab_L^{(n)} \in \Rset^d$, $d$ begin  the dimension of the measurement.  The
set $\F_L = \Big\{  f_{L,n} \Big\}_{n=0}^{2 \, L}$ forms a  basis for the signal
space of  dimension\footnote{Dimension refers  here to  the size  of the  set of
basis functions and not to $d$, that of the measurement.  The coefficients being
on  $\Rset^d$, when  $d  > 1$  the  notation  ``$\vect$'' is  thus  an abuse  of
writing.} $2 \, L + 1$, noted
\begin{equation}
\label{Eq:SignalSpace}
\E_L = \vect \F_L.
\end{equation}
The basis functions depend on geometrical parameters $(V , D , t_0)$ only, while
the $\ab_L^{(n)}$  depend on the  physical source (through $\mb^{(l)}$),  on the
angle $\beta$ and on the projection  matrix $\Pib$.  In a practical setting, $V$
is a parameter of the sensing process and  is known, while $D$ and $t_0$ must be
estimated.   This   will  not  be  discussed   here  (see~\cite{Blanpain1979PhD,
  Pepe2015generalization} for  an introductory discussion of  this problem), and
both $D$ and $t_0$ will be assumed to be known in the sequel. These results call
for a few additional comments:
\begin{itemize}
  \item    An     identical    expression     of    $\sb(u)$     was    obtained
    in~\cite{Pepe2015generalization}, starting  from another form  of multipolar
    expansion~\cite{Gonzalez1998multipole, Wikswoi1984multipoleexpansion} of the
    field satisfying the Laplace equation.
  \item  The basis  functions set  inferred  from a  dipolar source  field is  a
    particular  case  of $\F_L$  with  $L  =  1$.   These were  called  Anderson
    functions~\cite{Loane1976speed,                             Blanpain1979PhD,
      Baum1999DetectionObscuredTarget,  Ginzburg2002processing}  named after  J.
    E.    Anderson   who   was   apparently   the   first   to   exhibit   these
    functions~\cite{Anderson1949MAD}.
  \item Setting  the truncation  order $L$  has been so  far overlooked  in this
    paper. Although practical considerations preclude  to consider values of $L$
    much  larger than  $L  = 3$  (octupolar expansion)  as  the basis  functions
    decrease extremely fast  when $L$ increases, a brief  discussion is deferred
    to a future section.
\end{itemize}


\subsubsection{Nested multipolar subspaces and consequences}
\label{Sec:MS}

Restarting  from Eq.~\eqref{Eq:MagneticMultipoleTrajectory}  that expresses  the
field observed along the linear trajectory for a single pure $2^l-$polar source,
the following  $(l+2)$ dimensional subspace  to which  such a signal  belongs is
introduced:
\begin{equation}
\label{Eq:2lPolarSignalSpace}
\E^{(l)} = \vect \Big\{ f_{l,n} \Big\}_{n=0}^{l+1}  . 
\end{equation}
Dimension     $(l+2)$      comes     simply      from     the      fact     that
$\F^{(l)}~=~\Big\{f_{l,n}\Big\}_{n=0}^{l+1}$ forms  a free family  of functions.
From this definition, the following trivial inclusions may be established~:
\begin{equation*}
\E^{(l)} \subset \E_L,  \quad 1 \leqslant l \leqslant L, %
\qquad 
\E^{(1)} = \E_1.
\end{equation*}
Furthermore, it is easy to check that
\begin{equation}
\label{Eq:fN,n=fN+1,n+fN+1,n+2}
f_{L , n} =
f_{L + 1 , n} + f_{L + 1 , n+2} .  
\end{equation}
Imposing $n  \leqslant L$  insures that $f_{L,n}  \in \E^{(L)}$,  $f_{L+1,n} \in
\E^{(L+1)}$ and $f_{L+1,n+2}  \in \E^{(L+1)}$.  This shows  that $\E^{(L+1)}$ is
overlapping (or has a non-empty  intersection) with $\E^{(L)}$. Furthermore, the
condition $n \leqslant L$  implies that $\E^{(L)}$ is not a  subset of $\E^{(L +
  1)}$ (hopefully): actually,  using Eq.~\eqref{Eq:fN,n=fN+1,n+fN+1,n+2} and the
definition of the subspaces, it follows that for $L \geqslant 1$,
\begin{equation*}
\E_L \cap  \E^{(L + 1)} = \vect \Big\{
f_{L,n}
\Big\}_{n=0}^L \neq \emptyset.
\end{equation*}
This  simple   result  has  an  important   impact:  there  may  exist   a  pure
$2^{L+1}-$polar source leading to an observed  signal that may receive a sparser
representation  on  the  (shorter)  basis  obtained from  the  source  of  order
$L$. \modifications{For  example, it may  happen that, for a  purely quadrupolar
  source ($L = 2$), the signature  along the trajectory is entirely contained in
  the  dipolar signal  space; in  such a  configuration there  exists a  dipolar
  source that  generates exactly  the same measurement  along the  trajectory as
  that  obtained  for  the purely  quadrupole  source\footnote{\modifications{It
    should be noted  that an identification problem using a  measurement along a
    linear trajectory would  be ill-posed, since two  different physical sources
    (with different orders)  can lead to the same acquired  signal.}}.}  When it
comes  to  derive  detection  strategies  in  the  presence  of  noise,  sparser
representations  must be  preferred; actually,  it leads  to capture  all signal
energy on  a shorter number  of components while  the noise components  are kept
lower (see  later on).  Therefore, choosing  the right (the smallest)  value for
$L$ is an important issue, to be discussed later in the paper.

From now, we will  denote by $N$ the smallest order  allowing to fully represent
the signal  $\sb$ recorded along  the trajectory,  called in the  following {\it
  order of the signal}. Let us insist again on the fact that the magnetic source
is  not  necessarily multipolar  of  order  $N$,  but  necessarily of  order  $L
\geqslant N$.


\section{Magnetic Anomaly Detection formulation}
\label{Sec:MAD}


\subsection{The energy detector}
\label{Subsec:EnergyDetector}

Assuming additive noise, detection of a magnetic anomaly source is formulated as
a binary hypothesis testing problem
\begin{equation}
\label{Eq:binary_detection}
\left\{
\begin{array}{ll}
\H_0 \, : \, \xb \, = \, \nb       & \text{(absence of the source)}\\[1.5mm]
\H_1 \, : \, \xb \, = \, \sb + \nb & \text{(presence of the source)}
\end{array}
\right.
\raisetag{2\normalbaselineskip}
\end{equation}
In  practice all  observations are  sampled; let  $K$ be  the number  of samples
recorded  along  the  trajectory.    Considering  $d-$dimensional  sensors,  the
measurement   $\xb$,   the   signal   $\sb$   and   the   additive   observation
noise\footnote{Noise  accounts for  \enquote{uncontrolled} contributions  to the
measured  signal,  ranging  from  geomagnetism  phenomena  to  recording  device
eigen-noise and  motion related uncertainties~\cite{Ash1997Noise}.   Remind that
is is assumed that the Earth's field is subtracted from the measurement, so that
only  its  potential fluctuations  are  included  in  $\nb$, together  with  the
residues of the noise just evoked. } $\nb$, are all matrices in $\Rset^{d \times
  K}$.  It will be further assumed that the noise is uncorrelated to the signal,
and that it is a Gaussian white process with independent identically distributed
coordinates of variance $\sigma^2$. Thus, the noise probability density function
(pdf) reads:
\begin{equation}
\label{Eq:GaussianMatrixPDF}
p_{\nb}(\yb)  =  (2  \pi  \sigma^2)^{-\frac{K  d}{2}}  \,  \exp\left(  -\frac1{2
  \sigma^2} \Tr\left( \yb^\transp \yb \right) \right),
\end{equation}
where $\Tr$ stands for the trace operator, and $\yb^\transp$ is the transpose of
$\yb$.   Although  the assumptions  above  may  seem restrictive,  most  results
established in this  section and the following  will be valid in  a more general
setting, provided the noise correlation matrix is known or estimated in advance.
The discussion is thus postponed in appendix~\ref{App:ColoredNoise}.

Solving such a  binary test appears in many frameworks  such as radar detection,
communications or  geosciences. It  has received great  attention and  is widely
documented~\cite{Kay1998detection}. Whatever the  strategy developed for solving
this detection  problem, it involves  a comparison of the  (log)likelihood ratio
(LR)    to    some   threshold.     Exploiting    the    basis   exhibited    in
Eq.~\eqref{Eq:Signal}-\eqref{Eq:SignalBasis},   the    source-only   signal   is
expressed as
\begin{equation*}
\sb =  \Ab_N \Fb_N,
\end{equation*}
where $\Ab_N = \begin{bmatrix} \ab_N^{(0)} & \cdots & \ab_N^{(2N)} \end{bmatrix}
\in \Rset^{d  \times (2N+1)}$ and where  $\Fb_N \in \Rset^{(2N+1) \times  K}$ is
defined by  $[\Fb_N]_{ij} = f_{N,i}(j  \, \delta u)$ for  $0\leq i \leq  2N$ and
$0\leq j \leq K-1$;  $\delta u$ is the sampling step: the  $i-$th row of $\Fb_N$
contains $K$ values of $f_{N,i}$ sampled  on the linear trajectory of the sensor
device.   A given  trajectory and  a truncation  order $N$  completely specifies
$\Fb_N$. On the contrary, $\Ab_N$ appears  to be totally source (and $\beta$ and
sensor  axes  orientation)  dependent  and   appears  as  a  matrix  of  unknown
components.  In  the absence  of any  prior on  the probability  distribution of
$\Ab_N$, a generalized  log-likelihood ratio test (GLRT)~\cite{Kay1998detection}
is implemented:
\begin{equation*}
\max_{\Ab_N} \Lambda(\xb \, | \, \Ab_N  ) \: \equiv \: \log \left( \frac{p_{\nb}
  \left( \xb - \widehat{\Ab_N} \Fb_N \right) }{p_{\nb}(\xb)} \right) \: \test \:
\eta,
\end{equation*}
where  $\Lambda(\xb \,  |  \, \Ab_N  )$ denotes  the  log-likelihood ratio  when
$\Ab_N$ is known, and $\widehat{\Ab_N}$ is the maximum likelihood estimator.  In
the GLRT above, the left hand side is called {\it the receiver}. Receiver values
larger than the threshold $\eta$ lead to  decide that $\H_0$ is more likely than
$\H_1$  and  conversely; this  is  expressed  by  the double  inequality  symbol
$\displaystyle \test$.  Setting  $\eta$ depends on the  adopted strategy.  Here,
$\eta$ is chosen to maximize the probability of detection while constraining the
probability  of false  alarm (deciding  $\H_1$ while  $\H_0$ is  true) is  upper
bounded ({\it Neyman-Pearson}  strategy, see~\cite{Kay1998detection}).  As white
Gaussian  noise  is  assumed,  $\widehat{\Ab_N}$ matches  the  min-square  error
estimator of $A_N$:
\begin{equation*}
\widehat{\Ab_N}        =        \max_{\Ab_N}         \,        \Tr        \left(
(\xb-\Ab_N\Fb_N)(\xb-\Ab_N\Fb_N)^\transp \right).
\end{equation*}
Furthermore, setting the number of samples to  $K > 2N+1$ (this will be the case
in practice) we  insure that $\Fb_N$ has  full rank since $\F_N$  defined in the
previous  section forms  a free  family.  Consequently,  the Gram  matrix $\Fb_N
\Fb_N^\transp$  is  nonsingular   and  the  estimator  of   $A_N$  is  expressed
as\footnote{In  fact, the  results holds  considering the  Moore-Penrose inverse
even when  $\Fb_N \Fb_N^\transp$  is singular, but  the pseudo-inverse  does not
take expression~\eqref{Eq:MPInverse} anymore.}
\begin{equation}
\widehat{\Ab_N} = \xb \, \Fb_N^+,
\label{Eq:AMLE}
\end{equation}
with
\begin{equation}
\Fb_N^+ = \Fb_N^\transp \left( \Fb_N\Fb_N^\transp \right)^{-1}
\label{Eq:MPInverse}
\end{equation}
the  Moore-Penrose  pseudo-inverse  of  $\Fb_N$.   Plugging  the  expression  of
$\widehat{\Ab_N}$ in  the receiver  leads after some  algebra and  rejecting the
known quantities in the threshold to the expression of the test:
\begin{equation}
\label{Eq:Energy_test}
\Tr\left(\widehat{\Ab_N}\Fb_N\Fb_N^\transp\widehat{\Ab_N}^\transp   \right)   \,
\test \, \eta.
\end{equation}
Equivalently,   using  the   estimated  source   signal  $\widehat{\sb}   =
\widehat{\Ab_N}\Fb_N$:
\begin{equation*}
\Tr\left(\widehat{\sb} \, {\widehat{\sb}}^\transp \right) \test \eta.
\end{equation*}
This equation is nothing but the well-known energy detector.


\subsection{Practical issues and expected performance study}
\label{Subsec:IssuesPerformance}

Despite the simplicity  of the linear model  $\sb = \Ab_N \Fb_N$  assumed in the
previous   section,    some   difficulties   arise.    First,    the   test   in
Eq.~\eqref{Eq:Energy_test} requires evaluating  the Moore-Penrose pseudo-inverse
$\Fb_N^+$ of  $\Fb_N$.  This  may be  difficult or even  give rise  to numerical
instabilities.  Second, the statistical distribution of the receiver is required
in  order to  derive the  performances  of the  test, which  leads to  tractable
derivations if the $2N+1$ rows form an orthonormal basis.  Both arguments pledge
for the derivation of the test on some orthonormal basis.

Gram-Schmidt  orthonormalization  procedure  has  been applied  for  the  dipole
case~\cite{Blanpain1979PhD, Ginzburg2002processing} as well as for the multipole
one in~\cite{Pepe2015generalization}.   The discussion on  orthonormalization is
deferred to the next section.

For the  moment, let $\Gb_N$  be a given matrix  whose rows form  an orthonormal
basis; by misuse  of writing, extrapolating preceding notations,  let $\Ab_N$ be
the  vectors of  coefficients  of  the source  expressed  on  this basis.   Thus
Eqs.~\eqref{Eq:AMLE}-\eqref{Eq:Energy_test}  reduce  to  the projection  of  the
observation on $\Gb_N$, and to the energy detector
\begin{equation}
\widehat{\Ab_N} = \xb \, \Gb_N^\transp,
\qquad 
\left\| \widehat{\Ab_N} \right\|_F^2 \, \test \, \eta,
\label{Eq:GLLRTAO}
\end{equation}
where the  receiver $\Lambda \equiv  \| \Ab  \|_F^2 = \Tr\left(  \Ab \Ab^\transp
\right)$ is the squared Frobenius norm~\cite{Golub2013matrixcomputation}.

A  detector  such as  in  Eq~\eqref{Eq:GLLRTAO}  is  fully characterized  by  is
Receiver Operating Characteristics (ROC) (see~\cite{Kay1998detection}).  The ROC
is  parameterized  by $\eta$  and  expresses  the (conditional)  probability  of
detection $\Pd(\eta) = \Pr\left[  \Lambda > \eta \, \big| \,  \H_1 \right]$ as a
function of the (conditional) probability of false alarm $\Pfa(\eta) = \Pr\left[
  \Lambda > \eta \, \big|  \, \H_0 \right]$\footnote{A perfect detector exhibits
a ROC equal to {\it one} for  all $\eta \in \left( -\infty \, , +\infty\right)$,
while a random  detector satisfies $\Pd(\eta) = \Pfa(\eta), \:  \forall \eta \in
\left( -\infty  \, , \, +\infty  \right)$}. In order  to compute the ROC  of the
detector \eqref{Eq:GLLRTAO}, the statistical distributions of the receiver under
both $\H_1$ and  $\H_0$ must be computed.  For sake  of generality, let's assume
that the  detector is derived  with an arbitrary order\footnote{Remind  that the
\enquote{true} order of the source is generally unknown and even if it is known,
$\sb$ can  possibly be  represented by  a source  of a  lower order.   There are
actually  three different  orders involved  in practice:  the order  $L$ of  the
physical source (real or truncated), the  order $N$ of the signal recorded along
the trajectory, the order $M$ used at the receiver} $M$.  Then we get
\begin{equation*}
\widehat{\Ab_M} = \xb \, \Gb_M^\transp = k \, \sb_M \, + \nb_M,
\end{equation*}
where $k  \in \{0  \, ,  \, 1\}$  is a  factor associated  to $\H_0$  and $\H_1$
respectively, and
\begin{equation}
\sb_M = \sb \, \Gb_M^\transp \quad \mbox{and} \quad \nb_M = \nb \, \Gb_M^\transp
\label{Eq:ProjectedSignal}
\end{equation}
are the  projection of the signal  and the noise on  $\E_M$, respectively. Since
$\Gb_M^\transp$ has rank  $2 M + 1$ by construction,  the $d\times(2M+1)$ matrix
$\widehat{\Ab_M}$    can    be    shown    to    follow    a    matrix    normal
distribution~\cite[Eq.~2.3.10]{Gupta2018matrix}\footnote{$\Zb  \sim  \N_{p ,  q}
\big(  \zb ,  \Sigmab_p \otimes  \Sigmab_q \big)$  means that  the vectorization
$\vec\left(  \Zb^\transp   \right)  \in   \Rset^{p  q}$  whose   components  are
$\vec\left(  \Zb^\transp \right)_k  = \Zb_{i,j},  \: k  = (i-1)  p +  j$ ($\vec$
stacks the columns of a matrix) is Gaussian with mean $\vec(\mb)$ and covariance
matrix $\Sigmab_p \otimes \Sigmab_q$.}:
\begin{equation*}
\widehat{\Ab_M} \: \Big| \: \H_k \: \sim \: \N_{d  , 2 M + 1} \big( k \, \sb_M ,
\sigma^2 \Ib_d \otimes \underbrace{\Gb_M \Gb_M^\transp}_{\Ib_{2 M + 1}} \big).
\end{equation*}
It follows that ${\left\| \widehat{\Ab_M}  \right\|_F^2}/{\sigma^2}$ is a sum of
$d \, (2  M + 1)$ independent  squared standard normal random  variables, and is
distributed         according        to         a        chi-squared         law
$\chi^2_{\nu_M}\!(k\lambda_M)$~\cite{Johnson1995:v1,     Patnaik1949NonCentral},
with
\begin{equation}
\nu_M = d \, (2 M + 1),
\label{Eq:DegreeOfFreedom}
\end{equation}
degrees of freedom and noncentrality parameter $k \lambda_M$ with
\begin{equation}
\lambda_M = \dfrac{\| \sb_M \|_F^2}{\sigma^2}.
\label{Eq:NonCentralParameter}
\end{equation}
To  within  a  factor $1/{d  \,  K}$,  this  latter  is nothing  more  than  the
signal-to-noise ratio  (SNR) obtained after  the signal has been  projected onto
$\E_M$. It is worth noting that  this parameter is independent of the projection
matrix $\Pib$  characterizing the sensor  when it is  isometric\footnote{This is
for instance not the case for $d-$axes sensors with $d < 3$ (e.g., scalar) since
is this case the $3 \times 3$ matrix  $\Pib^\transp \Pib$ is of rank $d < 3$ and
thus  cannot be  the identity.   For  three-axes sensors  with orthogonal  axes,
$\Pib$ is  just a rotation  matrix and thus isometric:  in other words,  in this
case the result is  independent of the orientation of the  sensor.}.  The pdf of
the receiver is then obtained:
\begin{equation*}
\left\|  \widehat{\Ab_M}  \right\|_F^2  \:  \Big|   \:  \H_k  \sim  \sigma^2  \,
\chi^2_{\nu_M}\!(k \, \lambda_M).
\end{equation*}
The  expression of  the  distribution of  the receiver  above  allows to  easily
compute both false alarm and detection probabilities:
\begin{equation}
\label{Eq:PerfEta}
\left\{\begin{array}{lll}  \PfaM{M}(\eta) &  =  & \ccdf_{\chi^2_{\nu_M}}  \left(
\dfrac{\eta}{\sigma^2} \right) \\[4mm]
\PdM{M}(\eta)     &    =     &    \ccdf_{\chi^2_{\nu_M}\!(\lambda_M)}     \left(
\dfrac{\eta}{\sigma^2}\right)
\end{array}\right.,
\end{equation}
where  $\ccdf_{\chi^2_\nu(\lambda)}$  stands  for the  complementary  cumulative
density   function   (ccdf)   of  a   $\chi^2_\nu(\lambda)-$distributed   random
variable\footnote{The   noncentrality   parameter    is   omitted   when   zero,
$\chi^2_\nu(0) \equiv  \chi^2_\nu$.}.  The analytical  expression of the  ROC is
consequently given by:
\begin{equation}
\label{Eq:Pf_wrt_Pfa}
\PdM{M}\left(  \PfaM{M} \right)  = \ccdf_{\chi^2_{\nu_M}\!(\lambda_M)}  \compose
\ccdf^{-1}_{\chi^2_{\nu_M}} ( \PfaM{M} ),
\end{equation}
where $\compose$ is the composition  operator.  Remarks: The obtained ROC enjoys
the  following highly  desirable properties,  whose proofs  are deferred  in the
appendix~\ref{App:ROC}. The ROC is a  concave, increasing function of $\Pfa$; it
is  parameterized by  $\eta$,  with  fixed points  $(0,0)$  and  $(1,1)$ in  the
$\Pfa/\Pd$  plane   reached  for   $\eta  =  +\infty$   and  $\eta   =  -\infty$
respectively. The attainable  $\Pd$ for a given fixed value  of $\Pfa$ increases
when $\lambda_M$ increases or, equivalently, the  area under the ROC curve (AUC)
(defined by $\displaystyle  \AUC_M = \int_0^1 \PdM{M}\left(  \PfaM{M} \right) \,
d\PfaM{M}$)  is an  monotonic increasing  function  of $\lambda_M$:  this is  in
concordance with the fact that, for a given sample size $K$, higher SNR leads to
better detection performances or, for a  given SNR, larger sample size $K$ leads
to better detection performance.

All  the above  results  rely  on the  identification  of  an orthonormal  basis
(described  by the  matrix  $ \Gb_M$),  the  aim  of which  is  to avoid  matrix
inversion  and  to allow  analytical  derivation  of the  statistical  detection
performances.  Obtaining this  basis can lead to instabilities  in the numerical
orthonormalization process, particularly for a large set of basis functions.  In
order to circumvent these pitfalls, the calculation of an analytical orthonormal
basis is proposed in the next section.


\section{Analytical multipolar orthonormal basis functions}
\label{Sec:MOBF}

In this section, orthogonal polynomial theory is used to construct an analytical
orthonormal  basis   from  the   continuous-time  basis  $\F_N$   introduced  in
Eq.~\eqref{Eq:SignalBasis}.  It is shown that  the proposed method is equivalent
to a Gram-Schmidt  (GS) recursive orthonormalization procedure  while offering a
more direct approach.


\subsection{Orthonormalization}
\label{Subsec:Orthonormalization}

Recall that the  orthonormalization process is a change of  basis, replacing the
basis elements  $f_{N,n}$ by linear combinations  of them. Let $g_{N,n}$  be the
new  orthonormal basis  elements, their  (natural) inner  product satisfying  by
construction
\begin{equation}
\label{Eq:NaturalInnerProduct}
\int_\Rset g_{N,n}(u) \, g_{N,m}(u) \, du \, = \, \delta_{n,m},
\end{equation}
with  $\delta_{n,m}$ the  Kronecker symbol,  equal to  $1$ if  $n =  m$ and  $0$
otherwise~\cite{Szego1975OrthogonalPolynomials,              Arfhen2013MathPhys,
  Golub2013matrixcomputation}.   As the  $g_{N,n}$  are  linear combinations  of
$f_{N,n'}, \, n' = 0, \ldots , 2 \, N$, they satisfy
\begin{equation}
\label{Eq:gNnViaPNn}
g_{N,n}(u) = \frac{P_{N,n}(u)}{\left( 1 + u^2 \right)^{N + \frac32}},
\end{equation}
where  the  $P_{N,n}()$  are  polynomials  of degree  $n$.   The  natural  inner
product~\eqref{Eq:NaturalInnerProduct},  rewrites  as the  $w_N-$weighted  inner
product between polynomials $P_{N,n}$ and $P_{N,m}$:
\begin{equation}
\label{Eq:WeightedInnerProduct}
\int_\Rset \! P_{N,n}(u) \, P_{N,m}(u) \, w_N(u) \, du,
\end{equation}
where the weight function $w_N$ is given by
\begin{equation}
\label{Eq:WeightN}
w_N(u) =  (1+u^2)^{-2N-3}.
\end{equation}
As  a consequence,  the orthogonalization  problem  can be  reformulated in  the
framework          of         the          theory         of          orthogonal
polynomials~\cite{Szego1975OrthogonalPolynomials,
  Chihara2011OrthogonalPolynomials,  Arfhen2013MathPhys, Abramowitz1972handbook,
  Gradshteyn2015tableofintegrals}.   To  the  authors  knowledge,  no  classical
results    exist    for    the    specific   weight    function    defined    in
Eq.~\eqref{Eq:WeightN}. Therefore, most properties  and derivations are detailed
in  appendix~\ref{App:OrthonormalPolynomials},   along  with   technical  issues
related  to  the  present  problem.    Only  the  principal  results  and  their
consequences are discussed in this section.

By       relying      on       Rodrigues'      formula~\cite{Arfhen2013MathPhys}
or~\cite{Chihara2011OrthogonalPolynomials},   for   the   weight   function   in
Eq.~\eqref{Eq:WeightN}, we obtain the following series of polynomials satisfying
Eq.~\eqref{Eq:WeightedInnerProduct}:
\begin{equation}
\label{Eq:RodriguesN}
P_{N,n}(u) = c_{N,n} \, (1+u^2)^{2N+3} \, \dfrac{d^n}{du^n} (1+u^2)^{n-2N-3},
\end{equation}
where $c_{N,n}$  is the  normalization coefficient;  executing the  $n-$th order
differentiation                 of                 the                 composite
function~\cite[Th.~5.1.4]{Stanley2024enumerativecombinatorics},    after    some
algebra (see  appendix~\ref{App:OrthonormalPolynomials}) the polynomials  can be
written as
\begin{equation}
\label{Eq:PNn}
P_{N,n}(u)  =   c_{N,n}  \sum_{k=0}^{\left\lfloor   \dfrac{n}{2}  \right\rfloor}
d_{N,n, k} (1+u^2)^k \left( 2 u \right)^{n-2k},
\end{equation}
where $\lfloor \cdot \rfloor$ is the floor function and
\begin{equation}
\label{Eq:dNnk}
d_{N ,  n ,  k} = \dfrac{(-1)^{n-k}  \, n!  \,  (2N+2-k)!}{(2N+2-n)!  \,  k!  \,
  (n-2k)!}.
\end{equation}
Expressing the normalization constraint to  evaluate the constant term $c_{N,n}$
leads to solve
\begin{eqnarray*}
1 & = & \int_\Rset P_{N,n}(u)^2 \, w_N(u) \, du\\
& = & c_{N,n} \int_\Rset  P_{N,n}(u) \, \dfrac{d^n \, (1+u^2)^{n-2N-3}}{du^n} \,
du,
\end{eqnarray*}
where one factor  $P_{N,n}$ was replaced via the Rodrigues'  formula.  After $n$
successive   integrations   by  parts   and   some   calculations  detailed   in
appendix~\ref{App:OrthonormalPolynomials}, we get
\begin{equation}
\label{Eq:cNn}
c_{N,n}^2 = \dfrac{4^{2N+2-n} \, (4N+5-2n) \, \big( (2N+2-n)! \big)^2}{\pi \, n!
  \, (4N+5-n)! }.
\end{equation}
The following orthonormal basis $ \G_N  = \Big\{ g_{N,n} \Big\}_{n=0}^{2 N}$ for
the natural inner product is finally obtained:
\begin{equation}
\label{Eq:MOBF}
g_{N,n}(u) = \dfrac{P_{N,n}(u)}{(1+u^2)^{N+\frac32}},
\end{equation}
where $P_{N,n}$ is  given by Eqs.~\eqref{Eq:PNn}-\eqref{Eq:dNnk}-\eqref{Eq:cNn}.
$\G_N = \left\{ g_{N,n} \right\}_{n=0}^{2 N}$ will be called the {\it Multipolar
  Orthonormal Basis Functions} (MOBF).

Remarks:
\begin{itemize}
  \item  These functions  can  also  be expressed  in  terms  of Gegenbauer  (or
    ultraspherical) polynomials $C_n^{(2N-n+3)}$ of  degree $n$ and parameter $2
    N - n + 3$~\cite[Eq.~22.3.4]{Abramowitz1972handbook}:
  \begin{equation}
  \label{Eq:MOBF_Gegenbauer}
    g_{N,n}(u)   =  \frac{c_{N,n}   n!}{(2  N   +   2  -   n)!}   \left(   1+u^2
    \right)^{\frac{n-3}2   -  N}   C_n^{(2N-n+3)}\left(  \dfrac{u}{\sqrt{1+u^2}}
    \right).
  \end{equation}
  This alternate  formulation of the  basis $\Big\{ g_{N,n}  \Big\}_{n=0}^{2 N}$
  can be  useful for  numerical calculations,  since Gegenbauer  polynomials are
  widely implemented in many numerical softwares.
  \item {\it Gram-Schmidt (GS) equivalence}.   The set of orthogonal polynomials
    $\left\{ P_{N,n} \right\}_{n=0}^{2N}$  obtained previously coincides exactly
    with the  set obtained by applying  a GS orthogonalization procedure  to the
    $\F_N$  basis.   Although not  very  often  (explicitly) documented  in  the
    literature,  this is  a result  generic  to any  orthogonal polynomials,  as
    described  for instance  in~\cite{Szego1975OrthogonalPolynomials}.  A  brief
    proof is depicted in the appendix~\ref{Subsec:GSEquivalence}.  Nevertheless,
    despite its  recursive nature, the GS  approach does not (to  our knowledge)
    allow us to obtain an analytic expression of the basis.
\end{itemize}


\subsection{\modifications{Basis examples, $N = 2$ and $N = 3$}}
\label{Subsec:BasisDescription}

\begin{figure}[htbp]
\centering
\includegraphics[width=\linewidth]{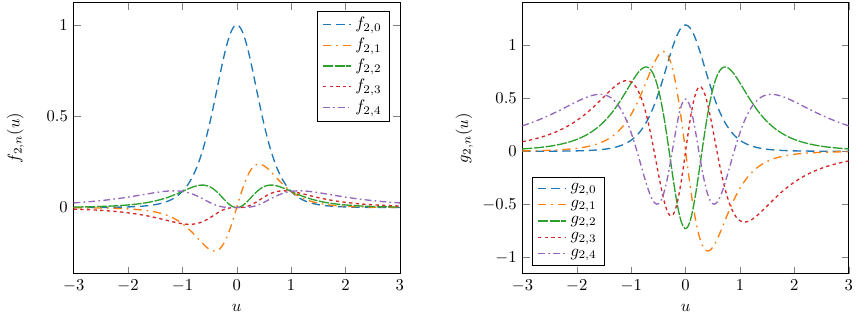}
\caption{The set of basis  functions $\F_2 = \Big\{f_{2,n}\Big\}_{n=0}^4$ (left)
  and set $\G_2 = \Big\{g_{2,n}\Big\}_{n=0}^4$ of MOBF (right).}
\label{Fig:base2}
\end{figure}

For the  purpose of  illustration, the case  of a  second order
\modifications{and the case of a third order} multipolar signal \modifications{are} used.
The  \modifications{sets} of  basis functions  $\F_2$ \modifications{and  $\F_3$
  are}      shown      in      Fig.~\ref{Fig:base2}-left      \modifications{and
  Fig.~\ref{Fig:base3}-left,  respectively};   the  \modifications{sets}  $\G_2$
\modifications{and $\G_3$} of MOBF  resulting from the orthogonalization process
applied    to   $\F_2$    \modifications{and   to    $\F_3$   are}    shown   in
Fig.~\ref{Fig:base2}-right     \modifications{and    Fig.~\ref{Fig:base3}-right,
  respectively}.

\modifications{
\begin{figure}[htbp]
\centering
\includegraphics[width=\linewidth]{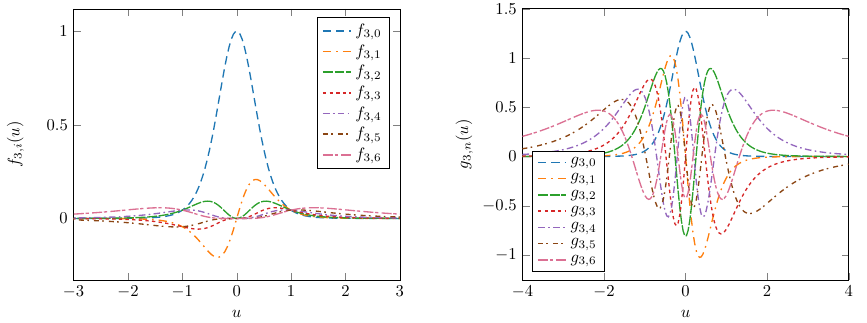}
\caption{
  \modifications{
    The set of basis  functions $\F_3 = \Big\{f_{3,n}\Big\}_{n=0}^6$ (left)
    and set $\G_3 = \Big\{g_{3,n}\Big\}_{n=0}^6$ of MOBF (right).
    }
}
\label{Fig:base3}
\end{figure}
}

\modifications{Figs.~\ref{Fig:base2}-right  and~\ref{Fig:base3}-right  evidence}
that $g_{2,n}$ \modifications{and $g_{3,n}$ have}  exactly $n$ roots. This holds
true for any order $N$ as the zeros  of $g_{N,n}$ coincide with the zeros of the
Gegenbauer  polynomials  on  $\left[  -1  \,  ,  \;  +1  \right]$  according  to
Eq.~\eqref{Eq:MOBF_Gegenbauer}; actually, it is  known that $C_n^{(2N-n+3)}$ has
exactly $n$ distinct zeros~\cite{Area2004zeros}.  This leads us to interpret the
projection onto the basis as a  decomposition into oscillating modes governed by
Gegenbauer polynomial roots.


\subsection{Numerical implementation of the basis}
\label{Subsec:NumericalStudy}

All  the developments  in the  previous  section assumed  continuous time  basis
functions.   However in  practice the  recorded  signals will  be sampled.   Let
$\big\{ u_k \big\}_{k = 1}^K$ define a regular sampling grid of $K$ samples over
a given integration window of width $R$  centered at $0$, i.e., $\left[ - R/2 \,
  ,  \, R/2  \right]$.  The  continuous time  natural inner  product defined  in
Eq.~\eqref{Eq:NaturalInnerProduct} will be approximated by the Riemann sum

\begin{equation*}
\int_\Rset  f(u) \,  g(u) \,  du \approx  \dfrac{R}{K-1} \sum_{k=1}^K  f(u_k) \,
g(u_k),
\end{equation*}
where  term on  the  right expresses  nothing but  the  discrete scalar  product
between sampled  functions $f$ and  $g$. As this  is only an  approximation, the
crucial  orthogonality  property necessary  to  derive  the simple  receiver  in
Eq.~\eqref{Eq:GLLRTAO} may be  lost. The purpose of this section  is to evaluate
the impact  of sampling parameters $K$,  $R$ (or specifically the  sampling step
$R/(K-1)$) and the multipolar order $N$ on the orthonormality of the basis. This
will  highlight that  the detection  performances will  not be  be significantly
altered by the sampling process.


\subsubsection{Impact of sampling  on orthogonality}
\label{Subsubsec:Orthogonality}

\begin{figure}[htbp]
\centering
\includegraphics[width=\linewidth]{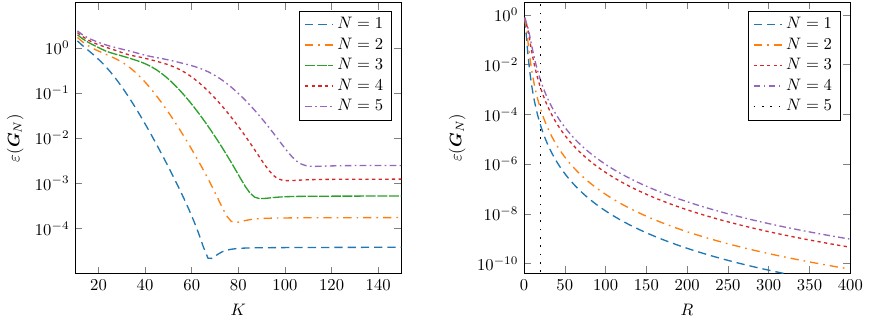}
\caption{Relative error  in terms of  Frobenius distance between Gram  matrix of
  $\Gb_N$ and identity  under different multipolar order $N$.   Left: w.r.t. $K$
  where $u$  is taken  in $\left[ -10  \, ,  \, 10 \right]$  ($R =  20$); Right:
  w.r.t. $R$ with constant sampling step $\dfrac{R}{K-1} = 2/100$.}
\label{Fig:gram_test_Rpas_constant}
\end{figure}

Let $\Gb_N$  be the discretization of  $\G_N$: the $n-$th row  of $\Gb_N$ counts
$K$ samples  of $g_{N,n}$,  normalized by  $\sqrt{\frac{R}{K-1}}$.  In  order to
measure   the   discrepancy   to    orthonormality,   the   following   relative
orthonormalization error of $\Gb_N$ is introduced:
\begin{equation*}
\varepsilon(\Gb_N) = \dfrac{\| \Gb_N \Gb_N^\transp - \Ib_{2 N + 1} \|_F}{\sqrt{2
    N + 1}},
\end{equation*}
where the denominator $  \sqrt{2 N + 1} = \| \Ib_{2 N  + 1} \|_F$, the Frobenius
norm of $\Ib_{2 N + 1}$ to which the Gram matrix $\Gb_N \Gb_N^\transp$ should be
equal to if it were  strictly orthonormal.  The behavior of $\varepsilon(\Gb_N)$
is represented (in  logarithmic scale) in Fig.~\ref{Fig:gram_test_Rpas_constant}
as a function of $K$ ($R$ fixed) or of $R$ ($K$ fixed).

Fig.~\ref{Fig:gram_test_Rpas_constant}-left                                  and
Fig.~\ref{Fig:gram_test_Rpas_constant}-right show that  increasing the number of
samples for  a fixed integration window,  or increasing window size  for a fixed
sampling step  increases the normality  of the basis.   This is expected,  as it
tends  to  cancel  out  the  effect of  discretization  and  finite  integration
window. The larger $K$ and the lower  $K/(R-1)$, the better: setting $K$ and $R$
will however be constrained by the signal recording physical characteristics.  A
simulation  for   realistic  parameters  (most   of  which  are   summarized  in
Table~\ref{Tab:params}) is described below, in  order to illustrate and evaluate
the effect of sampling the analytical orthonormal basis.
\begin{table}[htb]
\begin{minipage}{.36\linewidth}
  \renewcommand{\arraystretch}{1.25}
  \begin{tabular}{|c|c|}
  \hline 
  $V$ & $85~\si[per-mode=reciprocal,inter-unit-product=.]{\meter\per\second}$  \\
  \hline 
  $D$ & $100~\si{\meter}$  \\
  \hline 
  $K$ & $1001$~samples \\
  \hline 
  $R$ & $20, \qquad u \in [-10 \, , \, 10]$ \\
  \hline
  \end{tabular}
\end{minipage}\hfill
\begin{minipage}{.64\linewidth}
  \caption{Pseudo-operational parameters: rough approximation of parameters that
    can be used in operating conditions.}
  \label{Tab:params}
\end{minipage}
\end{table}
This choice of parameters will be referred to as \enquote{pseudo-operational} in
the   sequel,   corresponding  to   sample   an   airborne  sensor   signal   at
$42.5~\si{\hertz}$  during about  $23.53~\si{\second}$, giving  also $R/(K-1)  =
2/100$.   From   the  figures   above  for   these  \enquote{pseudo-operational}
parameters, orthogonalization error ranges from $3.95  . 10^{-5}$ for $N = 1$ to
$2.55 .  10^{-3}$ for $N = 5$, approximately, leading to a good approximation of
the scalar product by Riemann integration.

\begin{figure}[htbp]
\begin{minipage}{.5\linewidth}
  \includegraphics[width=.96\linewidth]{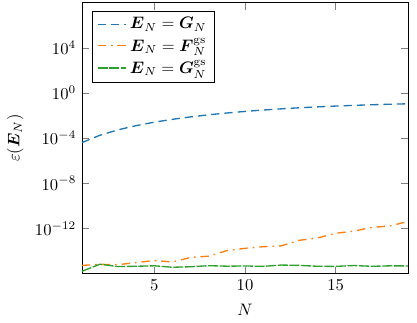}
\end{minipage}
\begin{minipage}{.5\linewidth}
  \caption{Frobenius  distance  $\varepsilon(\Eb_N)$   between  Gram  matrix  of
    $\Eb_N$  and  identity  w.r.t.   $N$, with  respectively  $\Eb_N  =  \Gb_N$,
    $\FNGS$, $\GNGS$ for parameters given in Table~\ref{Tab:params}.}
  \label{Fig:gram_test}
\end{minipage}
\end{figure}

Fig.~\ref{Fig:gram_test}  represents   the  evolution  of  the   deviation  from
orthonormality      ($\varepsilon(\cdot)$)     w.r.t.       $N$     for      the
\enquote{pseudo-operational} parameters  and for  three different  bases, namely
the sampled MOBF $\Gb_N$, then $\GNGS$  and $\FNGS$, the last two being obtained
after GS  orthonormalization of  $\Gb_N$ and  $\Fb_N$ respectively.   It appears
that  although $\Gb_N$  is  obtained  by truncating  on  $R$  then sampling  the
continuous analytical  orthonormal basis, it  remains fairly close to  a Stiefel
matrix\footnote{We refer  here to a matrix  $\Gb \in \Rset^{(2 N  +1) \times K}$
whose  transpose belongs  to Stiefel  manifold  on $\Rset$,  $\Gb \Gb^\transp  =
\Ib$~\cite{Gupta2018matrix,                 Muirhead1982MultivariateStatistical,
  Magnus2019matrix}} for  all $N$,  while $\varepsilon(\GNGS)$  and to  a lesser
extend $\varepsilon(\FNGS)$ are close to computational accuracy.

\subsubsection{Impact on detection performances}
\label{Subsubsec:SamplingAndROC}

\begin{figure}[htbp]
\begin{minipage}{.5\linewidth}
  \includegraphics[width=.96\linewidth]{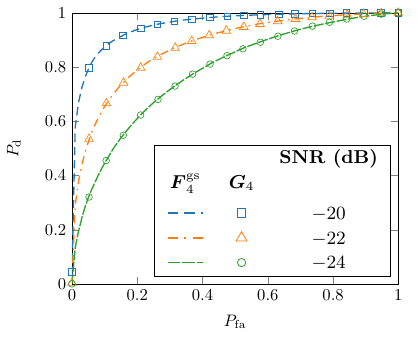}
\end{minipage}
\begin{minipage}{.5\linewidth}
  \caption{ROC  using  $\Fb_4^{\mathrm{gs}}$  and   $\Gb_4$  for  the  $4-$order
    receiver  for the  detection of  a $4-$order  anomaly, estimated  from Monte
    Carlo simulation with $10^5$ realizations of the noise, for various SNR.  }
  \label{Fig:ROC_test}
\end{minipage}
\end{figure}

The  impact  of   sampling  the  MOBF  is  to  introduce   some  discrepancy  to
orthonormality.   This latter  was shown  to  remain moderate  in the  preceding
section.   The purpose  of  this subsection  is  to assess  the  impact of  MOBF
sampling  on the  detection performances.   A set  of $10^5$  noisy signals  was
simulated from  the model  trajectory described in  Fig.~\ref{Fig:Geometry}, for
multipolar anomaly  of order  $L =  4$; A noise-free  signal was  generated from
equations
Eqs.~\eqref{Eq:MagneticField}-\eqref{Eq:partial-field}-\eqref{Eq:harmo-spher},
where the 24 harmonic coefficients $a_{lm}$,  $b_{lm}$ were drawn at random from
a standard  normal distribution, where  parameter $\beta$ was  randomly selected
from a uniform  distribution on $\left[ -\pi/2  \, , \, \pi/2  \right]$, and the
configuration   of  Table~\ref{Tab:params}   is   considered.    $d=3$  in   the
simulations, assuming  measurements with  a three  orthogonal axes  sensor, and,
without loss of generality, the axes are considered aligned with the trajectory.
The ROC  of the  energy detector  of Eq.~\eqref{Eq:GLLRTAO} taking  $M =  4$ and
using $\Fb_4^{\mathrm{gs}}$  or $\Gb_4$ are reported  on Fig.~\ref{Fig:ROC_test}
for various  signal-to-noise ratio (SNR)\footnote{In this  situation, looking at
the      expression      of       the      detection      performance      given
Eqs.~\eqref{Eq:DegreeOfFreedom}-\eqref{Eq:NonCentralParameter}-\eqref{Eq:PerfEta},
this  is   only  dependent  on   the  SNR   since  $\lambda_4  =   \frac{\|  \sb
  \|_F^2}{\sigma^2} $, and thus remain  unchanged whatever the $4-$order anomaly
and $\beta$.}.   Under the  assumption of  additive Gaussian  noise, the  SNR is
defined as
\begin{equation}
\SNR\text{(dB)}  =  10 \log_{10}  \left(  \dfrac{\|  \sb \|_F^2}{d  K  \sigma^2}
\right).
\label{Eq:SNR}
\end{equation}

The  striking  result  is that  sampling  the  MOBF  appears  to leave  the  ROC
unaltered, at least for the \enquote{pseudo-operational} parameters and SNR used
for the simulation.   This is also confirmed  by the evaluation of the  AUC as a
function of  SNR for  $\Fb_4^{\mathrm{gs}}$, $\Gb_4$  and $\Gb_4^{\mathrm{gs}}$,
reported in Table~\ref{Tab:classifier_feature_subsets_prec}.  Furthermore, these
result highlight that although applying GS orthogonalization on $\Gb_N$ improves
orthonormality,  it has  only  barely measurable  consequence  on the  detection
performances, and thus may be avoided  and spared.  To summarize, $\Gb_N$ allows
to reach good performances, while enjoying the highly valuable property of being
analytically known.

\begin{center}
\begin{threeparttable}[htb]
\caption{AUC (\%) of the detectors for different SNR.}
\label{Tab:classifier_feature_subsets_prec}
\small
\setlength\tabcolsep{0pt}
\begin{tabular*}{.7\linewidth}{@{\extracolsep{\fill}} l cc cc cc @{}}
\toprule
& \multicolumn{6}{c}{SNR (dB)} \\
\cmidrule{2-7}
 & $-25$ & $-24$ & $-23$ & $-22$ & $-21$  & $-20$\\
\midrule
$\Fb_4^{\mathrm{gs}}$ & $73.65$ & $78.12$ & $83.03$ & $87.68$ & $92.08$ & $95.59$\\
$\Gb_4$             & $73.59$ & $78.08$ & $83.07$ & $87.67$ & $92.01$ & $95.62$\\
$\Gb_4^{\mathrm{gs}}$ & $73.69$ & $78.17$ & $82.89$ & $87.76$ & $92.12$ & $95.67$\\
\bottomrule
\end{tabular*}
\end{threeparttable}
\end{center}

\


\section{Multipolar MAD in action}
\label{Sec:MMAD}

The choice of  order $M$ for the  detector (the number of columns  in $\Gb_M$ is
$2M+1$) has an  obvious impact on detection performance. Values  of $M$ that are
too low  lead to  the signal  being projected  onto a  low-dimensional subspace,
which introduces  approximation errors.  Conversely,  large values of  $M$, i.e.
much  larger  than  the  real  order   $N$  of  the  recorded  signal,  lead  to
contributions being  taken into account  that only involve projected  noise (see
Eq.~\eqref{Eq:GLLRTAO}).    As  detailed   in  section~\ref{Sec:MS},   a  subtle
distinction exists between  the multipolar order $L$ of the  source (which has a
physical reality) and the multipolar order $N \leqslant L$ of the signal. $N$ is
the minimal order  needed to represent the signal recorded  along the trajectory
(the lowest multipolar source order  leading to the same measurement).  Although
$M = N$ might seem to be the best choice, the examples below illustrate that the
distribution  of  signal energy  at  different  orders  has a  more  determining
influence on detection performances.


\subsection{Two experiments with a pure physical quadrupole}
\label{Subsec:TwoExperiments}

Two  scenarios  are simulated  for  discussion  of  the  impact of  model  order
selection on  detection performances. Both rely  on \enquote{pseudo-operational}
settings introduced in Section~\ref{Sec:MOBF} for  the signal recorder, $d = 3$,
assuming a  measure with a  three orthogonal axes  sensor, and, without  loss of
generality the sensor axes are assumed  aligned with the trajectory, whereas two
different  pure  quadrupole  sources  $\sce{1}$ and  $\sce{2}$  are  considered.
Sources                   are                   simulated                   from
Eq.~\eqref{Eq:harmo-spher}-\eqref{Eq:partial-field}-\eqref{Eq:MagneticField},
and scaled such that their energy are equal.  Their coefficient values are given
in     Table~\ref{Tab:coef_harmo}    tensor     expression    in     formulation
Eqs.~\eqref{Eq:MagneticMoments}-\eqref{Eq:MagneticMultipole}-\eqref{Eq:MagneticField}
(see~\cite{Wikswoi1984multipoleexpansion}):

\begin{threeparttable}[htb]
\caption{Harmonic  coefficients  and  trajectory   parameter  of  $\sce{1}$  and
  $\sce{2}$.}
\label{Tab:coef_harmo}
\small
\setlength\tabcolsep{0em}
\begin{tabular*}{\linewidth}{@{\extracolsep{\fill}}c c c c@{} c @{}c c c c@{\extracolsep{\fill}}}
\toprule
$\boldsymbol{\sce{1}}$  & \multicolumn{3}{c}{$m$}  & &  $\boldsymbol{\sce{2}}$ &
\multicolumn{3}{c}{$m$}\\
\cmidrule{2-4} \cmidrule{7-9}
 & $0$ & $1$ & $2$ &  & & $0$ & $1$ & $2$\\
\cmidrule{1-4}\cmidrule{6-9} $a_{2, m}$  & $-571.20$ & $109.49$ &  $~187.38$ & &
$a_{2, m}$ & $-40.99$ & $~~154.05$ & $-17.96$\\
$b_{2,  m}$  &  &  $191.18$  &  $-86.35$   &  &  $b_{2,  m}$  &  &  $-148.79$  &
$~~15.63$\\
\cmidrule{1-4}\cmidrule{6-9} $\beta$  & $-0.95~\si{\radian}$ &  & & &  $\beta$ &
$-0.57~\si{\radian}$\\ \bottomrule
\end{tabular*}
\end{threeparttable}\\

\

\noindent with corresponding magnetic tensors
\begin{equation*}
\mb^{(2)}_{\sce{1}} \! = \! 
\begin{bmatrix*}[r]
 44.9740  &  -13.7430  &    8.7129\\
-13.7430  &  -14.6709  &   15.2136\\
  8.7129  &   15.2136  &  -30.3031
\end{bmatrix*}\!\!,
\quad
\mb^{(2)}_{\sce{2}} \! = \!
\begin{bmatrix*}[r]
 -1.7706  &    2.4873  &   12.2588\\
  2.4873  &    3.9452  &  -11.8404\\
 12.2588  &  -11.8404  &   -2.1746
\end{bmatrix*}\!\!.
\end{equation*}

We projected  the signal generated  by these  parameters onto the  dipolar base,
checking  that  these  choices  of  values warrant  that  the  signal  lives  in
$\E^{(2)}$ and  not in the lower  dimensional space $\E_1$: both  signals are of
order $N = 2$, thus coinciding with the order of the physical source.

\begin{figure}[htbp]
\centering
\includegraphics[width=\linewidth]{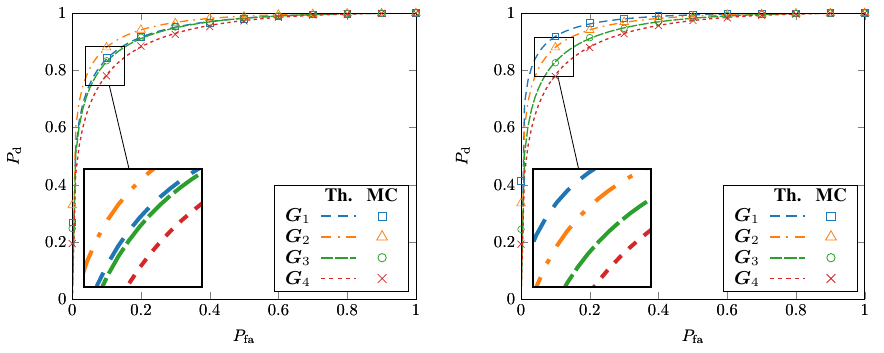}
\caption{ROC for $\sce{1}$ (left) and $\sce{2}$ (right) with $\SNR = -22$~dB for
  receiver~\eqref{Eq:GLLRTAO}   using  $\Gb_1,   \Gb_2,   \Gb_3$  and   $\Gb_4$.
  Monte-Carlo simulations are based on $10^4$ independent snapshots.}
\label{Fig:ROC_scenario}
\end{figure}

Theoretical  and experimental  (obtained by  Monte Carlo  simulations) ROCs  are
investigated for different multipolar orders $M \in \{1 \, , \, 2 \, , \, 3 \, ,
\,    4\}$    of   the    receiver,    at    a    fixed   $22$~dB    SNR    (see
Eq.~\eqref{Eq:SNR}). Results are reported on Fig.~\ref{Fig:ROC_scenario}.

It appears that  for both sources, the ROC deteriorates  as $M$ increases beyond
the    signal    order    $N    =     2$,    as    theoretically    proved    in
appendix~\ref{App:DecreasingROCwrtN}: in  such a situation, the  whole signal is
captured by the  receiver, and an increasing noise level  is projected in $\E_M$
as $M$ increases.  It  is worth noticing that the quadrupole  receiver ($M = 2$)
outperforms the  dipole ($M =  1$) receiver  in scenario $\sce{1}$,  whereas the
opposite is  \enquote{surprisingly} observed  for scenario  $\sce{2}$.  Elements
for discussion are introduced in the following subsection.


\subsection{Theoretical analysis of detection performances}
\label{Subsec:TheoreticalPerf}

Both the role  of the signal-to-noise ratio  after the receiver and  the role of
the energy distribution  on the different multipolar orders  are investigated in
the  sequel.  This  allows to  highlight the  role of  a proper  setting of  the
receiver  order   on  the   detection  performances,  and   provides  analytical
explanations on the experimental results from the preceding subsection.


\subsubsection{SNR based analysis}
\label{Subsubsec:SNRAnalysis}

Let  $\sb$ be  a $N-$order  multipolar  signal composed  of $K$  $d-$dimensional
samples and  $\alpha_N$ the proportion of  energy of this signal  projected onto
$\E_{N-1}$.  The  receiver projects  the observation  on the  $2M+1$ dimensional
space  $ \E_M$:  let $\sb_M$  and  $\nb_M$ the  projection of  $\sb$ and  $\nb$,
respectively (see Eq.~\eqref{Eq:ProjectedSignal}).  The signal-to-noise ratio of
the received signal is then written as
\begin{equation*}
\SNR_M = \dfrac{ \| \sb_M \|_F^2}{(2M+1) \, d \, K \sigma^2 }.
\end{equation*}
For $M = N$, $\sb_N = \sb$ and the receiver-projector does not induce any energy
loss.  Let $\alpha_N$ be defined by
\begin{equation*}
\alpha_N = \frac{\|  \sb_{N-1}\|_F^2}{\|\sb\|_F^2};
\end{equation*}
$\alpha_N$ characterizes the  latter loss when $M = N-1$,  and the deterioration
induced on the SNR is written as follows:
\begin{equation*}
\frac{\SNR_{N-1}}{\SNR_N} = \frac{\alpha_N \, (2 N + 1)}{2 N -1}.
\end{equation*}
For  $\alpha_N \geqslant  \dfrac{2N-1}{2N+1}$  the ratio  above  is larger  than
one. In  fact for $N =  2$, a projection  on $\E_1$ (perfect for  dipolar model)
will lead to a more favorable SNR than a projection onto $\E_2$ (associated to a
quadrupolar model)  as soon  as $\alpha_2 \geqslant  3/5$. Although  this result
makes it  possible to understand  the impact  of the choice  of $M$ on  the SNR,
detection performance  is governed  by the distribution  of test  statistics, of
which the  SNR gives  only a partial  view.  This is  confirmed by  the previous
simulations,  as both  scenario satisfy  $\alpha_2 \geqslant  3/5$ ($\alpha_2  =
0.747$ for $\sce{1}$  and $\alpha_2 = 0.941$ for $\sce{2}$),  but show different
behavior with regard to the choice of $M = 1$ or $M = 2$.


\subsubsection{ROC analysis}
\label{Subsubsec:ROCAnalysis}

For all subsequent  developments, a reference arbitrarily fixed  value $\Pfa$ of
the false alarm probability is considered, and  will not depend on the order $M$
chosen  for the  receiver.   The  derivation of  the  distribution  of the  test
statistics  in  Subsection~\ref{Subsec:IssuesPerformance}  allow  the  following
properties to be asserted:
\begin{enumerate}[label={$P_{\arabic*}$}]
  \item\label{Prop:DecreasingMHigherN}  For $M  \geqslant N$:  the noncentrality
    parameter in Eq.~\eqref{Eq:NonCentralParameter}  driving $\PdM{M}$ and given
    by  $\lambda_M   =  \dfrac{\|   \sb_M  \|_F^2}{\sigma^2}  =   \dfrac{\|  \sb
      \|_F^2}{\sigma^2}$  does  not  depend   on  $M$,  therefore  $\PdM{M}$  is
    decreasing     with      $M$.      The     proof     is      detailed     in
    appendix~\ref{App:DecreasingROCwrtN}.
  \item\label{Prop:AlphaCN} As  by construction $\alpha_N  \in [  0 \, ,  \, 1]$
    denotes the ratio of the SNR in the receiver projection space (see preceding
    subsection), $\PdM{N}  - \PdM{N-1}$ is  a function of $\alpha_N$,  and there
    exists      a       unique      \enquote{decision       critical      value}
    $\alpha_c(N,\Pfa,\lambda_N) \in [ 0 \, , \, 1]$ such that
  \begin{equation*}
  \alpha_N  \geqslant   \alpha_c\left(  N   ,  \Pfa   ,  \lambda_N   \right)  \:
  \Leftrightarrow \: \PdM{N-1} \geqslant \PdM{N}.
  \end{equation*}
  The elements of proof are the following:
  \begin{itemize}
    \item   If   $\alpha_N  =   0$,   then   $\PdM{N}  \geqslant   \PdM{N-1}   =
      \Pfa$. Actually, $\alpha_N  = 0 \: \Rightarrow \: \lambda_{N-1}  = 0 $ and
      $\PdM{N-1} =  \Pfa$ (see Eq.~\eqref{Eq:PerfEta}). The  result follows from
      the concavity of the ROC.
    \item If $\alpha_N  = 1$, then $\PdM{N} \leqslant \PdM{N-1}$.   In this case
      $\lambda_N = \lambda_{N-1}$, and the result  is a consequence of the decay
      of $\PdM{M}$ w.r.t. $M$.
    \item $\PdM{N} -  \PdM{N-1}$ is decreasing w.r.t.  $\alpha_N$:  In fact (see
      appendix~\ref{App:IncreasingROCwrtL}), $\PdM{N-1}$  is strictly increasing
      with the non  centrality parameter (proportional to the  SNR). This result
      holds as $\alpha_N$ is not involved in the expression of $\PdM{N}$.
  \end{itemize}
  \item\label{Prop:AlphaCN-1}  If  $\alpha_N  \geqslant   \alpha_c(N  ,  \Pfa  ,
    \lambda)$,  then $\exists!   \: \alpha_c\left(  N-1 ,  \Pfa ,  \lambda_{N-1}
    \right) \in [  0 \, , \,  1]$ such that $\alpha_{N-1}  = \dfrac{\| \sb_{N-2}
      \|_F^2}{\|  \sb_{N-1}  \|_F^2}  \geqslant  \alpha_c\left(  N-1  ,  \Pfa  ,
    \lambda_{N-1} \right) \: \Leftrightarrow  \: \PdM{N-2} \geqslant \PdM{N-1}$.
    This comes  by using the  same reasoning as  for property~\ref{Prop:AlphaCN}
    above.  Note  that this result  may be easily  extended by recursion  on the
    order $(N-k)$ of $\alpha_{(N-k)}$.
\end{enumerate}

Remarks:
\begin{itemize}
  \item Since  the AUC is  the integration  of $P_d(\Pfa)$, the  same conclusion
    about the existence  of an \enquote{AUC critical  values} $\alpha_c$ applies
    trivially to  the AUC, but  now such a critical  value no longer  depends on
    $\Pfa$.
  \item Properties~\ref{Prop:DecreasingMHigherN} to~\ref{Prop:AlphaCN-1} above
  highlight  that both  the noise  level  and the  way  in which  the energy  is
  projected  on different  multipolar orders  play  a strong  role in  detection
  performance.  This is  directly related  to the  importance of  the parameters
  $\sigma^2$ and  $\nu_M$ (degree of freedom)  in the definition of  $SNR_M$ and
  then  $\alpha_M$.   Despite  many  attempts,  no   analytical  expression  for
  $\alpha_c(M,\Pfa,\lambda_M), \: M \leqslant N$ has yet been found.
\end{itemize}

\begin{figure}[htbp]
\centering
\includegraphics[width=\linewidth]{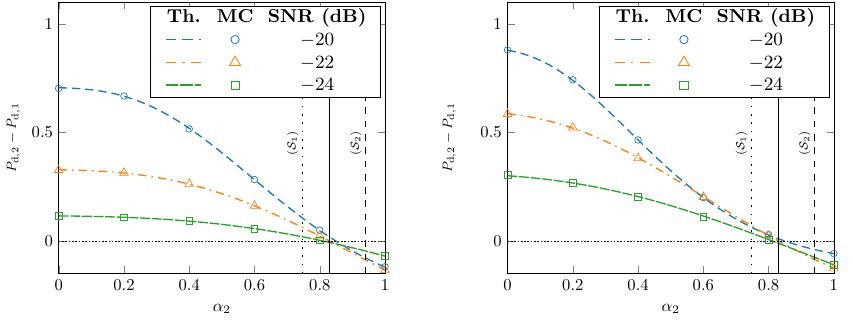}
\caption{$\PdM{2} - \PdM{1}$  w.r.t. $\alpha_2$ for $\Pfa =  10^{-3}$ (left) and
  $\Pfa = 10^{-2}$  (right). Dashed lines are analytically  computed, shapes are
  computed  with  Monte  Carlo  simulation with  $10^5$  snapshots.   The  black
  vertical line represents the crossover point between the curve at $-22$~dB and
  $0$. The vertical dotted lines represent  the proportion of dipolar energy for
  signals of scenarios $\sce{1}$ and $\sce{2}$.}
\label{Fig:pd2-pd1}
\end{figure}

Fig.~\ref{Fig:pd2-pd1}    illustrates     the    assertions     associated    to
properties~\ref{Prop:DecreasingMHigherN} to~\ref{Prop:AlphaCN-1}  above, for two
different fixed $\Pfa$  values, in the case $N  = 2$ and $\lambda_2 = d  \, K \,
\SNR$.    All  the   other  parameters   are   identical  to   those  from   the
\enquote{pseudo-operational}  setting  of  the   previous  section.   The  plots
represent the  variations of $\PdM{2}-\PdM{1}$  as a function of  $\alpha_2$ for
different  SNR values\footnote{The  theoretical  values are  comforted by  Monte
Carlo simulations which are obtained with  a toy signal \ $\sqrt{\alpha_2} \sb_1
+ \sqrt{1-\alpha_2} (\sb - \sb_1)$, with  $\sb$ the signal of scenario $\sce{1}$
and $\sb_1$ its projection into the dipolar signal space $\E_1$ \ ($\sb - \sb_1$
is  in its  orthogonal)}. Both  figures highlight  the existence  of a  decision
critical value $\alpha_c(2,\Pfa,\lambda_2)$ corresponding to the intersection of
$\PdM{2}-\PdM{1}$  with   the  horizontal   axis.   The  values   of  $\alpha_2$
corresponding  to  both scenarios  from  the  previous section  (represented  by
vertical  dotted lines)  appear to  be on  different sides  of $\alpha_c$,  thus
explaining the reason  why the best receiver is $\Gb_2$  for scenario $\sce{1}$,
whereas it is $\Gb_1$ for scenario $\sce{2}$ (see Fig.~\ref{Fig:ROC_scenario}).

\begin{figure}[htbp]
\begin{minipage}{.5\linewidth}
  \centering
  \includegraphics[width=.96\linewidth]{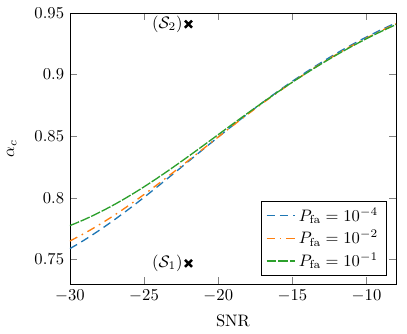}
\end{minipage}
\begin{minipage}{.5\linewidth}
  \caption{$\alpha_c\left(  2   ,  \Pfa   ,  \lambda_2   \right)$  w.    r.   t.
    $\SNR\text{(dB)}  = 10  \log_{10}\left( \dfrac{\lambda_2}{d  \, K}  \right)$
    (with the same $K$ as for  the previous examples), for various $\Pfa$. These
    curves are obtained  by searching numerically (e.g., by  dichotomy) the zero
    of $\alpha_2 \mapsto {\Pd}_{,2}(\Pfa) - {\Pd}_{,1}(\Pfa)$.  }
  \label{Fig:Critical}
\end{minipage}
\end{figure}

Finally, the  variations of $\alpha_c\left(  2 ,  \Pfa , \lambda_2  \right)$ are
shown in  Fig.~\ref{Fig:Critical}, as a function  of the SNR, for  several fixed
values  of $\Pfa$.   The decision  critical value  $\alpha_c$ appears  to depend
weakly on  $\Pfa$ in  the range  of interest $\Pfa  \in \left[  10^{-4} \,  , \,
  10^{-2} \right]$.

\modifications{
The same analysis is performed for $N=3$, leading to identical conclusion.  More
precisely, we are  considering an octupolar signal ($N=3$),  where $\alpha_3$ is
the energy proportion of this signal  projected onto $\E_{N-1} = \E_2$. Then, we
denote by $\alpha_N'$ the energy proportion of the projected signal $\sb_{N-1}$,
itself projected  onto $\E_{N-2} =  \E_1$.  In  other words, $\alpha_3'$  is the
proportion  of dipolar  energy  and $\alpha_3-\alpha_3'$  is  the proportion  of
quadrupolar energy when the dipolar  part is removed.  Thus $\alpha_3-\alpha_3'$
is  the proportion  of  signal energy  in the  dipolar  signal space  orthogonal
complement into the quadrupolar signal space.

The  detection  probability  of  detection  is  a  function  of  $\alpha_3$  and
$\alpha_3'$.  Fig.~\ref{Fig:Zones}  describes the receiver  order $\displaystyle
M_\opt= \argmax_{M \in \Nset^*} {\Pd}_{,M}$  with the best detection probability
in           the           $(\alpha_3'          ,           \alpha_3-\alpha_3')$
plane\footnote{\modifications{Necessarily  $0  \leqslant  \alpha_3  -  \alpha_3'
  \leqslant 1-\alpha_3' \leqslant  1$.}} for fixed both  false alarm probability
and SNR.  As for the  quadrupolar case, $M_\opt$  depends on $\Pfa$,  $\SNR$ and
$\alpha_3$, but also on $\alpha_3'$.

\begin{figure}[htbp]
\begin{minipage}{.5\linewidth}
  \centering
  \includegraphics[width=.96\linewidth]{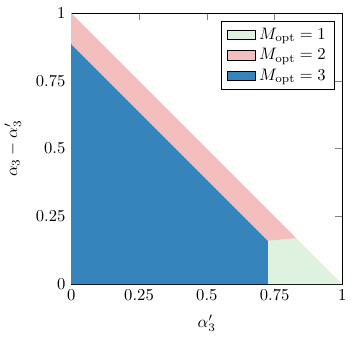}
\end{minipage}
\begin{minipage}{.5\linewidth}
  \caption{ \modifications{Case $N = 3$: theoretical order of the best receiver,
      $\displaystyle M_\opt = \argmax_{M \in \Nset^*} {\Pd}_{,M}$, as a function
      of the proportion of dipole energy and the proportion of quadrupole energy
      in the  orthogonal of the dipole.   The false alarm probability  is set to
      $\Pfa = 10^{-2}$ and the the signal-to-noise ratio to $\SNR = -22$~dB.}}
  \label{Fig:Zones}
\end{minipage}
\end{figure}

The shape of  this figure is hardly surprising: (i)  for a predominantly dipolar
signal, the  dipolar receiver prevails  (green area,  bottom right); (ii)  for a
predominantly quadrupolar  signal orthogonal  to the  dipolar signal  space, the
quadrupolar  receiver prevails  (top left);  (iii)  finally, for  a signal  that
exists mainly  in the signal  space orthogonal  to the quadrupolar  signal space
(dipole and quadrupole combined), the octupolar receiver prevails (bottom left).

Although not  shown here,  it can  be observed that  the different  zones weakly
depend  on $\Pfa$.  In addition,  the shape  of the  different zones  is similar
regardless of  the SNR;  the area of  the $M_\opt  = 1$ and  $M_\opt =  2$ zones
decrease as $\SNR$ increases.
}

At this stage it is important to emphasize that this section highlights the fact
that  the choice  of  the receiver  order  $M$ does  not  necessarily match  the
multipolar order  of the  source, nor does  it match the  order of  the measured
signal on a  given trajectory. Setting $M$ correctly has  a measurable impact on
the detection performances, as we have  just shown.  An operational approach for
selecting $M$ is therefore desirable; this is discussed in the next section.


\section{Choosing the order using information criteria}
\label{Sec:Criteria}

Selecting the  signal order  $N$ has  already been  described as  being possibly
handled using  Akaike information criterion  (AIC)~\cite{Akaike1974new} approach
in~\cite{Pepe2015generalization}. In this section  we extend this perspective to
the estimation  of the receiver order  $M$ since it  is $M$ that appears  in the
detection  scheme.  Furthermore, different  criteria  are  also considered.   An
original strategy  based on a  binary decision  approach is then  introduced and
evaluated for the estimation of $M$.


\subsection{Model order selection: introduction}
\label{Subsec:InformationCriteria}

The derivation  of model-order  selection criteria  pertains to  quite different
approaches,  clearly   presented  and  discussed   in~\cite{Stoica2004}.   While
Bayesian  information   criterion  (BIC)~\cite{Schwartz1978estimating}  approach
relies on maximum a posteriori model  parameters estimation, AIC is derived from
statistical  information theoretical  concepts.   It is  noteworthy that  coding
theory approach was investigated by Rissanen~\cite{Rissanen1978modeling} and led
to the minimal description length (MDL) principle for model order selection.  It
turns out that  for our model, MDL  and BIC lead to identical  expression of the
criterion; thus, MDL will not be mentioned anymore in the sequel.  Despite their
conceptual differences, all  these criteria end up being written  as follows for
our problem:
\begin{equation}
\label{Eq:criterion}
\left\{ \begin{array}{lll}
\Crit(M)  & =  & \dfrac{\big\|  \widehat{\Ab_M}  \big\|_F^2 }{\sigma^2}  -
c(M)\\[4mm]
\Mc & = & \displaystyle \argmax_{M \in \Nset^*} \, \Crit(M)
\end{array}\right.
\end{equation}
In the  expression above, $\Crit  \in\{\AIC, \BIC\}$ and  $ \Mc$ stands  for the
selected  model-order when  the criterion  $\Crit$  is used.   The variable  $c$
depends on the chosen criterion: for instance, $c(M)  = 4 \, d \, M$ for AIC and
$c(M)  = 2  \, d  \,  M \ln(dK)$  for  BIC\footnote{All the  constant terms  are
removed,  as having  no influence  on the  criteria. Moreover,  the opposite  is
considered  in the  literature,  and  the minimum  is  searched  instead of  the
maximum, which is obviously completely equivalent}.  Some generalizations of AIC
were  proposed in  the  literature (see~\cite{Stoica2004}),  but  all enter  the
generic formula  given in Eq.~\eqref{Eq:criterion}  for different values  of the
variable $c(M)$.  Therefore the methodology proposed below remains unchanged and
the focus will be solely on AIC and BIC.

One may note that $\Crit(M) $ expresses  a trade-off between the likelihood of a
model  of order  $M$, which  increases with  $M$, and  a penalization  term that
depends on  the chosen criterion,  whose role  is to prevent  overfitting.  This
approach can  also be interpreted here  as considering the GLRT  not only w.r.t.
$\Ab_M$, but also  w.r.t.  the order $M$, where the  latter part is penalization
by $c(M)$.


\subsection{Binary choice framework}
\label{Subsec:CritreriaOrderSelection}

Since the criterion $\Crit$ in Eq.~\eqref{Eq:criterion} is a random variable, so
is $\Mc$.   An exhaustive study  would require determining the  probability mass
function  (pmf) of  $\Mc$, which  is  difficult and  outside the  scope of  this
paper. Instead, we propose to calculate the pdf of
\begin{eqnarray*}
\Delta\Crit(M,m) & = & \Crit(M)  - \Crit(M-m)\\[2mm]
&  = &  \dfrac{\big\|  \widehat{\Ab_M} \big\|_F^2  - \big\|  \widehat{\Ab_{M-m}}
  \big\|_F^2}{\sigma^2} - \delta c(M,m),
\end{eqnarray*}
with
\begin{equation*}
\delta c(M,m) = c(M) - c(M-m),
\end{equation*}
whose sign  determines the choice $\Mc  = M$ or $\Mc  = M-m$ in a  binary choice
framework. Without loss of generality\footnote{Note that $\Delta\Crit(M,M+m) = -
\Delta\Crit(M+m,M)$.}, we consider $M \geqslant 2$ and $0 < m < M$.

For  any  value of  $M$,  $\big\|  \widehat{\Ab_M}  \big\|_F^2 =  \left\|  \xb_M
\right\|_F^2$,  where  $\xb_M$ is  the  projection  of  $\xb$ onto  $\E_M$.   By
construction  (see preceding  sections),  and for  any value  of  $m$, $\xb_M  -
\xb_{M-m}$ is orthogonal to $\xb_{M-m}$, and by Pythagorean theorem, we get
\begin{eqnarray*}
\Delta\Crit(M,m)
&  =  &  \dfrac{\left\|  \xb_M  -  \xb_{M-m}  \right\|_F^2}{\sigma^2}  -  \delta
c(M,m)\\[2mm]
& = &  \dfrac{\left\| \xb \, \gb_{M,m}^\transp  \right\|_F^2}{\sigma^2} - \delta
c(M,m),
\end{eqnarray*}
where $\gb_{M,m} \in  \Rset^{2 \, m \times K}$  is a matrix whose $2  \, m$ rows
provide  an  orthonormal  basis  of  the orthogonal  of  $\E_{M-m}$  in  $\E_M$.
Following  similar arguments  as above,  we  obtain $\left\|  \sb_M -  \sb_{M-m}
\right\|_F^2  = \left\|  \sb_M \right\|_F^2  - \left\|  \sb_{M-m} \right\|_F^2$.
Thus, using  notations and  results from  Sec.~\ref{Subsec:IssuesPerformance} we
conclude that
\begin{equation*}
\Delta\Crit(M,m) \, \Big| \, \H_k \: \sim  \: - \, \delta c(M,m) + \chi^2_{2 d
  m}( k \, \delta \lambda(M,m) )
\end{equation*}
with
\begin{equation*}
\delta\lambda(M,m) = \lambda_M - \lambda_{M-m}.
\end{equation*}


\subsection{Quadrupolar vs dipolar receiver model}
\label{Subsec:QuadrupolarVsDipolar}

The same scenarios  $\sce{1}$ and $\sce{2}$ as previously  are considered again,
both    involving   quadrupolar    sources.    As    it   was    emphasized   in
Sec.~\ref{Sec:MMAD}, depending on the proportion $\alpha_2$ of dipolar energy in
the  quadrupolar  signal,  selecting  $M  = 1$  may  lead  to  better  detection
performances than selecting $M  = 2$. In the sequel, the  analysis above is thus
applied  for  $M  = N$  and  $m  =  1$  (actually  $N  = 2$  for  $\sce{1}$  and
$\sce{2}$). Then  $(\lambda_N - \lambda_{N-1}) =  (1-\alpha_N)\lambda_N$ and the
decision rule is
\begin{equation*}
\left\{ \begin{array}{llll}
\mbox{if} & \Delta\Crit(N,1) > 0 & \mbox{choose} & \Mc = N\\[2mm]
\mbox{if} & \Delta\Crit(N,1) \leqslant 0 & \mbox{choose} & \Mc = N-1
\end{array}\right.
\end{equation*} 
The probability of the choice in this   binary setting is given by
\begin{equation}
\label{Eq:ProbaCritBinary}
\Pr[ \Mc  = N \,  | \,  \H_k] = \ccdf_{\chi^2_{2  d}( k \,  (1-\alpha_N) \,
  \lambda_N )} \left( \delta c(N,1) \right).
\end{equation}
In particular,  the criterion-based  binary choice leads  to choose  more likely
$\Mc = N$ rather than $\Mc = N-1$  if the above probability is larger that $.5$.
This   last   formulation   is    particularly   enlightening:   As   shown   in
appendix~\ref{App:IncreasingROCwrtL},  $\forall \,  \delta, \:  \alpha_N \mapsto
\ccdf_{\chi^2_{2  d}( (1-\alpha_N)  \, \lambda_N  )} \left(  \delta \right)$  is
decreasing  with $\alpha_N$  so that  there is  at most  one value  (function of
$\delta$)   of   $\alpha_N$  at   which   the   function  equals   $1/2$.    Let
$\alpha_{c,\Crit} \equiv  \alpha_{c,\Crit}\left( N , \lambda_N  \right)$ be this
\enquote{probability critical value}. If it exists, $\alpha_{c,\Crit}$ satisfies
\begin{equation*}
\ccdf_{\chi^2_{2d}(( 1-\alpha_{c,\Crit})  \, \lambda_N  )} \left(  \delta c(N,1)
\right) = \frac12.
\end{equation*}
This  last  equation  implies  that  $\delta   c(N,1)$  is  the  median  of  the
distribution  $\chi^2_{2d}(( 1  - \alpha_{c,\Crit}  ) \,  \lambda_N)$, and  this
implicitly defines  $\alpha_{c,\Crit}$, although this  does not not allow  for a
simple analytical  expression.  In  summary, focusing on  $\H_1$ (the  target is
present in the observation)
\begin{equation*}
\Mc  =  N  \:  \mbox{  is  more  likely  chosen  than  }  \:  \Mc  =  N-1  \quad
\Leftrightarrow \quad \alpha_N > \alpha_{c,\Crit}.
\end{equation*}

An alternate  point of view  consists in analyzing the  choice in terms  of {\it
  statistical   average}:    $\Mc   =   N$    is   chosen   in    average   when
$\Esp{\Delta\Crit(N,1)} > 0$.  From the expression  of the mean on a non-central
chi-squared   distribution~\cite[Sec.~1.3]{Muirhead1982MultivariateStatistical},
focusing   on    $\H_1$,   $\Esp{\Delta\Crit(N,1)}$   cancels   out    for   the
\enquote{average critical value}
\begin{equation*}
\overline{\alpha}_{c,\Crit}\left(  N,\lambda_N  \right)  = \left(  1  -  \dfrac{
  \delta c(N,1) - 2 \, d }{\lambda_N} \right)_{\!+},
\end{equation*}
where $( \cdot )_+ = \max( \cdot , 0 )$.  In summary, focusing on $\H_1$
\begin{equation*}
\Mc = N  \: \mbox{ in average rather  than } \: \Mc =  N-1 \quad \Leftrightarrow
\quad \alpha_N > \overline{\alpha}_{c,\Crit}.
\end{equation*}

\begin{figure}[htbp]
\begin{minipage}{.5\linewidth}
\centering
\includegraphics[width=.96\linewidth]{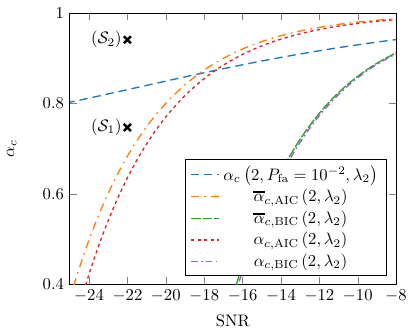}
\end{minipage}
\begin{minipage}{.5\linewidth}
  \caption{$N = 2$: Theoretical critical value $\alpha_c \equiv \alpha_c\left( 2
    ,  \Pfa =  10^{-2} ,  \lambda_2  \right)$ and,  $\alpha_{c,\Crit}\left( 2  ,
    \lambda_2  \right)$  and  $\overline{\alpha}_{c,\Crit}\left( 2  ,  \lambda_2
    \right)$, obtained  from the  criteria $\Crit  = \AIC$  and $\Crit  = \BIC$,
    w.r.t. $\SNR = \dfrac{\lambda_2}{d \, K}$.}
  \label{Fig:CriticalCriteria}
\end{minipage}
\end{figure}

The   values   obtained  numerically   for   the   probability  critical   value
$\alpha_{c,\Crit}(2 ,  \lambda_2)$ and those  computed for the  average critical
value    $\overline\alpha_{c,\Crit}(2    ,    \lambda_2)$   are    plotted    on
Fig.~\ref{Fig:CriticalCriteria}  for  various  SNR together  with  the  decision
critical value $\alpha_c\left( 2 , \Pfa = 10^{-2} , \lambda_2 \right)$.

This figure leads to the following observations:
\begin{itemize}
  \item The critical values appear to increase  with the SNR: at high SNR value,
    the (relatively  small) contribution  of quadrupolar  terms in  the receiver
    still contributes to detection performances,  whereas they are hidden in the
    noise at low SNR.
  \item  $\overline\alpha_{c,\Crit}$  is   close  to  $\alpha_{c,\Crit}$.   Both
    critical values lead  to conclude that the criteria  favor dipolar receiver,
    while  the \enquote{ground-truth}  $\alpha_c$ indicates  that a  quadrupolar
    receiver has better performances in scenario $\sce{1}$.  Although this seems
    disappointing, it  should be  remembered that the  criteria are  analyzed in
    terms of probabilities or statistical averages,  which does not rule out the
    quadrupolar receiver being selected for certain scenario realizations.
  \item  Other  choices  expanding   the  AIC  (see~\cite{Stoica2004})  lead  to
    different expressions of the variable $c$, that may be better adapted to our
    scenarios, and are deferred to future studies.   In any case we must keep in
    mind that all the criteria provide only asymptotic or approximate estimators
    of the order of the model.  The important results here lie in the ability of
    the information criterion  based proposed approach to  explain the existence
    of a critical value of the energy proportion between quadrupolar and dipolar
    signal  components,  beyond  which  we observe  a  discrepancy  between  the
    theoretical order  of the source  ($N$) and the  best order of  the receiver
    ($\Mc$). This result  is consistent with the SNR based  analysis proposed in
    Sec.~\ref{Sec:MMAD}.
\end{itemize}

In the subsection to come, numerous  simulations are introduced to highlight the
impact of the estimated receiver order on the detection performances.


\modifications{
\subsection{Octupolar vs quadrupolar vs dipolar receiver model}
\label{Subsec:QuadrupolarVsDipolar}

The same  analysis is performed in  the context $N=3$, for  ``high'' and ``low''
SNRs. In  Figs.~\ref{Fig:AIC_1vs2vs3} we consider  the AIC criterion  in context
$M=1$ versus $M=2$  versus $M=3$ and describe  $M^\star_{\mathrm{aic}}$ the most
probably  selected  order by  the  criterion.  The  dashed line  represents  the
delimitation  of   the  optimal  orders  $M_\opt=1$   versus  $M_\opt=2$  versus
$M_\opt=3$.

\begin{figure}[htbp]
\centering
\includegraphics[width=\linewidth]{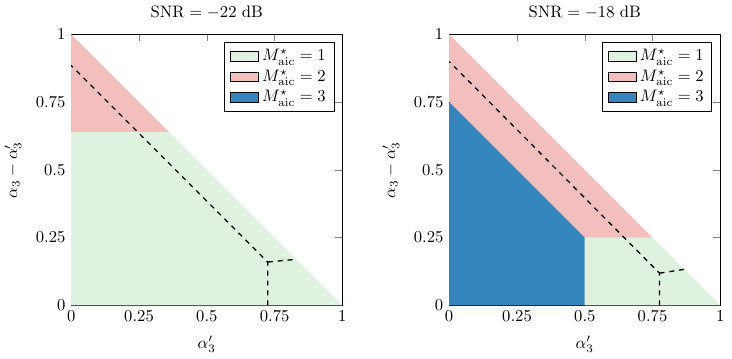}
\caption{\modifications{$N=3$:  multipolar   order  $M^\star_{\mathrm{aic}}$  of
    higher average  from the criteria $\Crit  = \AIC$ in a  context $M=1$ versus
    $M=2$  versus  $M=3$.   $\Pfa  =  10^{-2}$  with  $\SNR=-22$~dB  (left)  and
    $\SNR=-18$~dB  (right). The  optimal order  delimitations are  given by  the
    dashed lines.}}
\label{Fig:AIC_1vs2vs3}
\end{figure}

As in the  case of the octupole,  AIC remains imperfect: we  would have expected
that  the  AIC  selected  order  coincides  as  closely  as  possible  with  the
theoretical  orders.  At  $\SNR  =  -22$~dB the  deviation  between the  optimal
selection  order  and the  most  probable  AIC  selected order  is  significant;
moreover,  the probability  that  AIC selects  $M=3$ is  never  the highest  one
whatever the distribution of the signal  energy across the subspaces. At $\SNR =
-18$~dB, we observe that the blue area, corresponding to the highest probability
to select $M=3$, reappears and that  the deviation between the ``AIC selection''
and the optimum selection drops down.
}


\subsection{Information Criteria: detection performance}
\label{Subsec:CriteriaPerformance}

\begin{figure}[htbp]
\centering
\includegraphics[width=\linewidth]{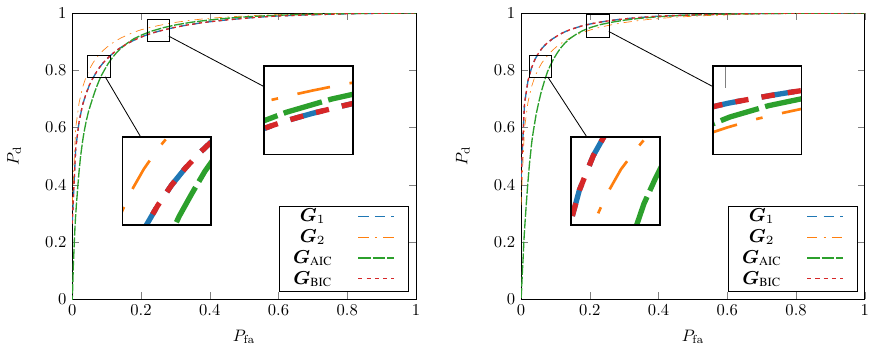}
\caption{ROC for $\sce{1}$ (left) and $\sce{2}$ (right) with $\SNR = -22$~dB for
  receiver~\eqref{Eq:GLLRTAO}  using  $\Gb_1,   \Gb_2,  \Gb_{\AIC},  \Gb_{\BIC}$
  computed by Monte Carlo simulation with $10^5$ snapshots.}
\label{Fig:roc-aic}
\end{figure}

An  important issue  lies  in  evaluating the  detection  performances when  the
receiver order is estimated (for each  new snapshot) by an information theoretic
approach as described above.  A  theoretical analysis remains however difficult,
as it would require to identify the pdf of $\Mc$, this latter being in turn very
difficult\footnote{For the criterion $\Crit$,  we have $$\displaystyle \Pr\left[
  \left.   \left\| \widehat{A_{\Mc}}  \right\|_F^2  > \eta  \,  \right| \,  \H_k
  \right]  =  \sum_{M  \geqslant  1}  \Pr\left[  \left.   \left\|  \widehat{A_N}
  \right\|_F^2  >  \eta   \,  \right|  \,  \Mc   =  N  \,  ,   \,  \H_k  \right]
\ \Pr\left[\left. \Mc = M \, \right|  \, \H_k \right];$$ the difficulties lie in
the determination of the probability law of $\Mc$, but also on the fact that the
$\widehat{A_N}$ and $\Mc$  are dependent (except for $N =  1$), i.e., the factor
of the mass function  of $\Mc$ is not ${\Pfa}_N$ under  $\H_0$ and not ${\Pd}_N$
under $\H_1$.}.  Instead, Monte-Carlo simulations  are conducted to evaluate the
ROCs.   Both scenarios  $\sce{1}$ and  $\sce{2}$  are considered  again and  the
resulting ROCs are drawn on Fig.~\ref{Fig:roc-aic}, for each considered receiver
$\Gb_{\Crit}$ and compared to $\Gb_1$ and $\Gb_2$.

\begin{figure}[htbp]
\centering
\includegraphics[width=\linewidth]{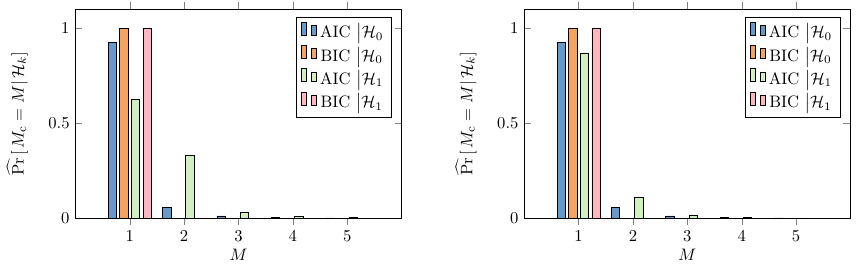}
\caption{Estimated distribution of $\Mc$ under $\H_k$ for $k = 0 , 1$ and $\Crit
  =  \AIC, \BIC$  on scenario  $\sce{1}$ (left)  and scenario  $\sce{2}$ (right)
  computed  by  Monte Carlo  simulation  with  $10^5$  snapshots under  $\SNR  =
  -22$~dB.}
\label{Fig:histo}
\end{figure}

The first observation is that $\Gb_\BIC$  always matches $\Gb_1$ as predicted in
Fig.~\ref{Fig:CriticalCriteria}: BIC criterion seems to be too conservative (the
penalty term is too high) in our  setting and for the chosen SNR.  This analysis
is supported by Fig.~\ref{Fig:histo} where  the histograms of the selected model
order are  plotted, showing that the  BIC criterion leads systematically  to the
selection $\Mbic = 1$.  Quantitatively, the probability to choose $\Mbic = 1$ is
quite accurately approximated by Eq.~\eqref{Eq:ProbaCritBinary}, $\Pr[ \Mbic = 1
  \, |  \, \H_1] \approx 1  - \ccdf_{\chi^2_{2 d}( (1-\alpha_2)  \, \lambda_2 )}
\left( 2 \, d \, \ln( d K) \right) \approx 1$, which supports the observation.

At the  opposite, we expected that  using AIC-based order selection  $\Maic$, if
not outperforms the fixed quadrupolar and dipolar receivers $\Gb_1$ and $\Gb_2$,
would at  least lead to  a trade  off between both  by making a  balance between
selecting $\Maic =  1$ and $\Maic = 2$,  as it would be the case  if we restrict
the  order  selection  to  be  binary   and  under  the  (false)  assumption  of
independence between  $\left\| \widehat{A_N} \right\|_F^2$ and  $\Maic$. Indeed,
under $\H_1$,  Fig.~\ref{Fig:histo}-left compared  to Fig.~\ref{Fig:histo}-right
shows that the AIC  criterion leads to the selection $\Mc =  2$ with fairly high
probability when  the energy proportion  of the dipolar  contribution decreases.
Again, the probability to choose $\Maic = 2$ is quite accurately approximated by
Eq.~\eqref{Eq:ProbaCritBinary},  $\Pr[  \Maic   =  2  \,  |   \,  \H_1]  \approx
\ccdf_{\chi^2_{2 d}( (1-\alpha_2) \, \lambda_2 )} \left( 4 \, d \right)$, and as
already commented, the  latter is increasing w.r.t.,  $\alpha_2$.  However, this
effect leads  to the expected trade  off $\Gb_1$ vs $\Gb_2$  only for relatively
high $\Pfa$,  while any of  the two fixed receiver  perform better for  the more
interesting regime of low $\Pfa$.

For both criteria, we must keep in mind that in selection model problems we deal
with data generated by the model, corrupted by (additive) noise. This not always
the case here  since, by essence of the  decision problem, we do not  know if we
are under $\H_1$ (context of standard model selection) or $\H_0$ (fit of a model
from noise  only). A criterion  can thus  be efficient to  fit a model  when the
signal is present in data, while having negative impact in detection since a fit
can increase  the rate  of false  alarm in the  absence of  the signal.   A more
quantitative explanation is given in appendix~\ref{App:PerfCriteria}.

As a  conclusion, beyond  the study of  the accuracy of  the order  selection in
itself, a study of the theoretical performance as a function of the penalization
term $c$ may lead to its \enquote{optimal} expression.


\section{Conclusion, discussion and perspective}
\label{Sec:Conclusion}

In this study,  we first set out  a new construction of  an analytical multipole
basis to describe the magnetic field produced by a fixed general source recorded
by a  $d-$axes sensor  along a  linear trajectory.   This construction  takes up
previous results obtained in~\cite{Pepe2015generalization}, in a more direct and
simplified approach while using the same assumption parameters. Next, the vector
subspaces of the various multipolar orders were analyzed.  Relationships between
these  observation subspaces  were  exposed and  shown to  lead  to cases  where
multipolar sources can live on observation subspaces associated with lower-order
multipolar  sources.  The  major consequence  is the  possible mismatch  between
source  truncation  order  and  optimal receiver  order.   To  enable  practical
analysis of the properties of the receiver, an analytical orthogonal basis based
on  the theory  of continuous  orthogonal polynomials  has been  proposed.  This
avoids recourse  to the orthogonalization procedure  or Moore-Penrose inversion,
which  are known  to  generate  higher computing  costs  and possible  numerical
instabilities.  Although these developments were carried out for continuous-time
signals, their extension  to discrete signal recordings, even  of high multipole
order, proved to maintain very  close detection performance.  Secondly, in order
to  design a  complete  detection system,  some insights  were  given about  the
receiver optimal order through a simple scenario of pure quadrupolar source.  It
was then shown the existence of a  threshold in the dipolar energy proportion of
the signal  measured along  a trajectory  that plays an  important role  when it
comes  to choosing  such  an  optimal receiver  order:  although  the signal  is
quadrupolar,  above  this  threshold  a lower-order  receiver  leads  to  better
performance.  In the  realistic context ,since the signal to  detect is unknown,
so is the distribution of the signal energy across the orders and this threshold
cannot  help for  the receiver  order choice.   To overcome  this difficulty,  a
simplified approach  involving classical criteria such  as AIC and BIC  has been
proposed for order selection in the detection context. A study in the example of
the quadrupolar source reveals that a  not too conservative criterion such a AIC
is to be preferred.  However, it  leads to performance not fully satisfactory as
making a  trade off between the  performance of the dipolar  and quadrupolar one
only for ``large'' probabilities of false  alarm, while in many applications the
regime  of interest  is more  likely that  of ``small''  probabilities of  false
alarm.  The  strategy to satisfactorily  choose the order remain  thus partially
open.

Other rather  important issues  are also  left for  further studies.   A problem
concerns the white noise assumption in the detection model.  Although no results
are  presented here,  the method  to handle  colored additive  noise is  briefly
sketched  in  the  appendix~\ref{App:ColoredNoise}.   Nevertheless,  simulations
should be thoroughly investigated to match realistic operational situations.

Moreover, the estimation of $D$ and  $t_0$ have not been treated.  Estimation of
$t_0$ is not  a problem, since under operational conditions,  the algorithm will
be executed at each time step over a sliding window.  However, the estimation of
$D$  remains  open.   The  brute-force   search  for  maximum  likelihood  on  a
finite-size     grid     has     been     proposed     in~\cite{Blanpain1979PhD,
  Pepe2015generalization}.   In the  near  future, this  optimization should  be
reconsidered in the light of the powerful development of optimization algorithms
(non-linear                and/or               non-convex                and/or
geometric)~\cite{BenTal2001ModermConvexOptimization,
  BenTal2023ModermConvexOptimization,            Nesterov2018ConvexOptimization,
  Stein2024NonlinearOptimization,  Rubinov2003LagrangeTypeNonConvexOptimization,
  Theobald2024RealAlbrebraicGeometryOptimization}.

Finally, efforts have
focused on determining  the \enquote{optimal} multipole order  for the receiver.
Although this  has some  analogy with  the search  for a  sparse representation,
another approach would be  to restrict the multipole order of  the receiver to a
certain  maximum  order,  and  then  identify a  sparse  representation  of  the
observation on this basis. Such an  approach is common in machine learning-based
modeling  and  relies  on  regularization  methods.   This  approach  should  be
investigated with a view to future MAD development.


\begin{appendices}


\section{Increasing and concavity properties of the ROC}
\label{App:ROC}

As usual, the ROC curve  $\PdM{M}\left(\PfaM{M}\right)$ is increasing from point
$(0,0)$ to $(1,1)$ and concave.

In the sequel, let us omit the subscript $M$ for the ease of reading, and let us
denote  $\vartheta  =  \frac{\eta}{\sigma^2}$.   Let  us  recall  that  $\Pfa  =
\ccdf_{\chi_\nu^2}(\vartheta), \:  \Pd = \ccdf_{\chi_\nu^2(\lambda)}(\vartheta),
\: \vartheta  \geqslant 0$ given Eq.~\eqref{Eq:PerfEta},  where $\ccdf_\zeta$ is
the  ccdf  of  random  variable $\zeta$  (respectively  central  and  noncentral
chi-squared variable here).

From the properties  of a ccdf, it  is immediate that, as well  known, that when
$\vartheta \to  0$ one has  $(\Pfa,\Pd) \to (1,1)$,  and, at the  opposite, when
$\vartheta \to +\infty$ one has $(\Pfa,\Pd) \to (0,0)$.

The increasing property is obvious also, from
\begin{equation*}
\dfrac{d\Pd}{d\Pfa}        =         \dfrac{\dfrac{\partial        \Pd}{\partial
    \vartheta}}{\dfrac{\partial    \Pfa}{\partial    \vartheta}}   =    \dfrac{-
  \pdf_{\chi^2_\nu(\lambda)}(\vartheta)}{-   \pdf_{\chi^2_{\nu}}(\vartheta)}  \:
\geqslant 0,
\end{equation*}
where  $\pdf_\zeta$  denote   the  probability  density  function   (pdf)  of  a
$\zeta-$distributed random variable.

In the same line, one can see that
\begin{equation}
\label{Eq:d2Pd/dPfa2}
\dfrac{d^2\Pd}{d\Pfa^2}   =    \dfrac1{\left(   \dfrac{\partial   \Pfa}{\partial
    \vartheta}\right)^3}  \left[  \dfrac{\partial^2  \Pd}{\partial  \vartheta^2}
  \dfrac{\partial \Pfa}{\partial  \vartheta} -  \dfrac{\partial^2 \Pfa}{\partial
    \vartheta^2} \dfrac{\partial \Pd}{\partial \vartheta} \right].
\end{equation}
Note  first that  $\left( \dfrac{\partial  \Pfa}{\partial \vartheta}\right)^3  =
\left( - \pdf_{\chi_{\nu}^2}(\vartheta)  \right)^3 \: \leqslant \: 0$.  To show the
concavity, one  has to prove  that the term in  square brackets is  positive. An
easy  way to  check this  is  to start  from the  expression  of the  ccdf of  a
noncentral chi-squared  distribution as  a mixture  of central  chi-squared ccdf
with  weights as  Poisson  probabilities  with rate  $\lambda/2$~\cite[Corollary
  1.3.5]{Muirhead1982MultivariateStatistical} or~\cite{Patnaik1949NonCentral},
\begin{equation}
\label{Eq:NoncentralChi2_PoissonMixture}
\ccdf_{\chi_{\nu}^2(\lambda)}    =    \sum_{k     \geqslant    0}    \dfrac{e^{-
    \frac{\lambda}2} \lambda^k}{2^k \, k!} \: \ccdf_{\chi_{\nu + 2 k}^2},
\end{equation}
which, in a sense, \enquote{decouple} $\nu$ and $\lambda$. Then, from
\begin{equation*}
\pdf_{\chi_\nu^2}(\vartheta)            =           \frac{\vartheta^{\frac{\nu}2-1}
  e^{-\frac{\vartheta}2}}{2^{\frac{\nu}2}  \Gamma\left(   \frac{\nu}2  \right)},
\quad  f'_{\chi_\nu^2}(\vartheta) =  \frac{\left( \nu  - 2  - \vartheta  \right)
  \vartheta^{\frac{\nu}2-2}            e^{-\frac{\vartheta}2}}{2^{\frac{\nu}2+1}
  \Gamma\left( \frac{\nu}2 \right)},
\end{equation*}
with $f'$ the derivative of $f$, we obtain
\begin{equation}
\label{Eq:fpkf-fpfk}
\displaystyle f'_{\chi_{\nu+2k}^2}(\vartheta)  \, \pdf_{\chi_{\nu}^2}(\vartheta) \,
- \, f'_{\chi_{\nu}^2}(\vartheta) \, \pdf_{\chi_{\nu+2k}^2}(\vartheta)
\displaystyle    =    \,    \frac{2    \,   k    \,    \vartheta^{\nu+k-3}    \,
  e^{-\vartheta}}{2^{\nu+k+1}  \Gamma\left(   \frac{\nu}2  \right)  \Gamma\left(
  \frac{\nu}2 + k \right)} \: \geqslant \: 0.
\end{equation}
Plugging      this      result      into      Eq.~\eqref{Eq:d2Pd/dPfa2}      via
Eq.~\eqref{Eq:NoncentralChi2_PoissonMixture}  allows to  close  the proof  since
that, the  term in parenthesis of  Eq.~\eqref{Eq:d2Pd/dPfa2} is then the  sum of
positive terms, and thus is indeed positive.


\section{Increasing property of the ROC w.r.t. $\lambda_M$}
\label{App:IncreasingROCwrtL}

The result in intuitively obvious, but let us prove it analytically.

A relatively  simple approach  lies again  in the  expression of  the ccdf  of a
noncentral chi-squared  distribution as a  mixture of central  chi-squared given
Eq.~\eqref{Eq:NoncentralChi2_PoissonMixture}.     Now,   differentiating    this
expression w.r.t. $\lambda$ gives
\begin{equation*}
\dfrac{\partial}{\partial  \lambda}  \ccdf_{\chi_{\nu}^2(\lambda)} =  -  \frac12
\sum_{k  \geqslant  0} \dfrac{e^{-  \frac{\lambda}2}  \lambda^k}{2^k  \, k!}  \:
\ccdf_{\chi_{\nu  +  2   k}^2}  +  \frac12  \sum_{k   \geqslant  1}  \dfrac{e^{-
    \frac{\lambda}2} \lambda^{k-1}}{2^{k-1}  \, (k-1)!} \: \ccdf_{\chi_{\nu  + 2
    k}^2},
\end{equation*}
which gives, after a change of indices in the second sum
\begin{equation}
\label{Eq:PartialFbar/PartialLambda}
\dfrac{\partial}{\partial   \lambda}  \ccdf_{\chi_{\nu}^2(\lambda)}   =  \frac12
\sum_{k \geqslant  0} \dfrac{e^{- \frac{\lambda}2} \lambda^k}{2^k  \, k!} \left(
\ccdf_{\chi_{\nu + 2 k + 2}^2} \!\! - \ccdf_{\chi_{\nu + 2 k}^2} \right).
\end{equation}
Then, from  $\displaystyle \ccdf_{\chi_a^2}(\vartheta) = 1  - \frac{\gamma\left(
  \frac{a}2 \,  , \, \frac{\vartheta}2\right)}{\Gamma\left(  \frac{a}2 \right)}$
of the ccdf  of a central chi-squares  law~\cite{Johnson1995:v1}, where $\gamma$
is the  incomplete gamma  function, the recurrence  relation $\gamma(a+1,v)  = a
\gamma(a,v) - v^a e^{-v}$~\cite[Eq.~6.5.22]{Abramowitz1972handbook} gives
\begin{equation}
\label{Eq:RecurenceFbarnuk}
\ccdf_{\chi^2_{a+2}}(\vartheta)  =  \ccdf_{\chi^2_a}(\vartheta) +  \dfrac{\left(
  \frac{\vartheta}2 \right)^{\frac{a}2} \, e^{- \frac{\vartheta}2}}{\Gamma\left(
  \frac{a}2 + 1 \right)}.
\end{equation}
We thus obtain, with $a = \nu+2k$,
\begin{equation*}
\ccdf_{\chi_{\nu +2 + 2 k}^2}(\vartheta) - \ccdf_{\chi_{\nu + 2 k}^2}(\vartheta)
=    \dfrac{\left(    \frac{\vartheta}2    \right)^{\frac{\nu}2+k}    \,    e^{-
    \frac{\vartheta}2}}{\Gamma\left( \frac{\nu}2  + k + 1  \right)} \, \geqslant
\, 0,
\end{equation*}
with  equality  if   and  only  if  or  $\vartheta  =   0$,  or  $\vartheta  \to
+\infty$.  Putting this  result in  Eq.~\eqref{Eq:PartialFbar/PartialLambda}, we
can conclude that,
\begin{equation*}
\dfrac{\partial}{\partial \lambda} \ccdf_{\chi_{\nu}^2(\lambda)} \, > \, 0 \quad
\mbox{on} \quad \left( 0 \, , \, +\infty \right),
\end{equation*}
so    that,    from     $\Pd(\Pfa)    =    \ccdf_{\chi^2_\nu(\lambda)}    \left(
\ccdf_{\chi^2_\nu}^{-1}(\Pfa)  \right)$ given  expression~\eqref{Eq:Pf_wrt_Pfa},
for a given $\Pfa \not\in \{ 0 \, , \, 1 \}$, the probability of detection $\Pd$
in strictly increasing  w.r.t., $\lambda$: as expected, the ROC  are improved as
$\lambda$ increases.  In particular, $\AUC$  is thus also a  strictly increasing
function w.r.t. $\lambda$.


\section{Source of order $N$: The performance is decreasing w.r.t. $M \geqslant N$}
\label{App:DecreasingROCwrtN}

We   start  again   from  $\Pd(\Pfa)   =  \ccdf_{\chi^2_\nu(\lambda)}   \compose
\ccdf_{\chi^2_\nu}^{-1}(\Pfa)$   given   expression~\eqref{Eq:Pf_wrt_Pfa}   and,
again, a key lies in Eq.~\eqref{Eq:NoncentralChi2_PoissonMixture}, that gives
\begin{equation*}
\Pd(p)  = \sum_{k  \geqslant 0}  \dfrac{e^{-\frac{\lambda}2} \lambda^k}{2^k  k!}
\ccdf_{\chi^2_{\nu + 2 k}} \compose \ccdf_{\chi^2_\nu}^{-1}(p).
\end{equation*}
Now,    if   for    any   $k$,    $\ccdf_{\chi^2_{\nu   +    2   k}}    \compose
\ccdf_{\chi^2_\nu}^{-1}  \geqslant  \ccdf_{\chi^2_{\nu  +  2 +  2  k}}  \compose
\ccdf_{\chi^2_{\nu+2}}^{-1}$, since  $\nu_{M+1} = \nu_M  + 2 \,  d$ ($d =  1$ or
$3$), we would have that for a given $\Pfa$ and constant $\lambda_M$ w.r.t. $M$,
$\Pd$ is decreasing w.r.t. $M$.

Let  us  then   prove  that,  indeed,  $\ccdf_{\chi^2_{\nu  +   2  k}}  \compose
\ccdf_{\chi^2_\nu}^{-1}  \geqslant  \ccdf_{\chi^2_{\nu  +  2 +  2  k}}  \compose
\ccdf_{\chi^2_{\nu+2}}^{-1}$.
First, denoting, for any $k$,
\begin{equation}
\label{Eq:UPfa}
\vartheta   =   \ccdf_{\chi^2_\nu}^{-1}(p)   ,   \quad   d_k   =   \dfrac{\left(
  \frac{\vartheta}2                 \right)^{\frac{\nu}2+k}                 e^{-
    \frac{\vartheta}2}}{\Gamma\left( \frac{\nu}2 + k + 1 \right)},
\end{equation}
to simplify the notations, we have,
\begin{eqnarray*}
\ccdf_{\chi^2_{\nu   +  2   k}}  \compose   \ccdf_{\chi^2_\nu}^{-1}(p)  &   =  &
\ccdf_{\chi^2_{\nu + 2 k}}(\vartheta) \\[2mm]
& = & \ccdf_{\chi^2_{\nu + 2 + 2 k}}(\vartheta) - d_k\\[2mm]
&  = &  \ccdf_{\chi^2_{\nu  +  2 +  2  k}} \compose  \ccdf_{\chi^2_{\nu+2}}^{-1}
\compose \ccdf_{\chi^2_{\nu + 2}}(\vartheta) - d_k,
\end{eqnarray*}
where we have used  Eq.~\eqref{Eq:RecurenceFbarnuk} with $a = \nu +  2 k$ in the
second line. Thus, from Eqs.~\eqref{Eq:RecurenceFbarnuk}-\eqref{Eq:UPfa} with $a
= \nu, \: k = 0$ we obtain
\begin{equation}
\ccdf_{\chi^2_{\nu  +   2  k}}  \compose  \ccdf_{\chi^2_\nu}^{-1}(p)   \:  =  \:
\ccdf_{\chi^2_{\nu +  2 + 2  k}} \compose \ccdf_{\chi^2_{\nu+2}}^{-1}\left(  p +
d_0 \right) - d_k.
\label{Eq:Diff1}
\end{equation}
Then,   noting    that   Eq.~\eqref{Eq:fpkf-fpfk}   say   nothing    more   that
$\ccdf_{\chi^2_{\nu + 2 k}} \compose \ccdf_{\chi^2_\nu}^{-1}$ is concave for any
$\nu, k$, replacing $\nu$ by $\nu+2$,
\begin{eqnarray*}
\ccdf_{\chi^2_{\nu +  2 + 2  k}} \compose \ccdf_{\chi^2_{\nu+2}}^{-1}\left(  p +
d_0    \right)    -    \ccdf_{\chi^2_{\nu    +    2    +    2    k}}    \compose
\ccdf_{\chi^2_{\nu+2}}^{-1}(p)
&  \geqslant  &  d_0  \,  \left(   \ccdf_{\chi^2_{\nu  +  2  +  2  k}}  \compose
\ccdf_{\chi^2_{\nu+2}}^{-1} \right)' \! \left( p + d_0 \right)\\[2mm]
&   =   &   d_0   \:   \dfrac{\pdf_{\chi^2_{\nu   +   2   +   2   k}}   \compose
  \ccdf_{\chi^2_{\nu+2}}^{-1} \left(  p +  d_0 \right)}{\pdf_{\chi^2_{\nu  + 2}}
  \compose \ccdf_{\chi^2_{\nu+2}}^{-1} \left( p + d_0 \right)}\\[2mm]
& =  & d_0  \left( \dfrac{\ccdf_{\chi^2_{\nu+2}}^{-1} \left(  p +  d_0 \right)}2
\right)^{\!   k}  \dfrac{\Gamma\left(   \frac{\nu}2  +  1  \right)}{\Gamma\left(
  \frac{\nu}2 + k + 1 \right)},
\end{eqnarray*}
where  we have  used in  the fourth  line the  expression $\pdf_{\chi^2_a}(v)  =
\dfrac{v^{\frac{a}2-1}  e^{- \frac{v}2}}{2^{\frac{a}2}  \Gamma\left( \frac{a}{2}
  \right)}$ of the pdf of a chi-squared law with $a  = \nu + 2 + k$ and $a = \nu
+   2$,   respectively.    Thus,   with   $a   =   \nu$   gives   $\vartheta   =
\ccdf_{\chi^2_{\nu+2}}^{-1} \left(  p + d_0 \right)$,  so that the last  term in
the previous inequality is nothing but $d_k$, i.e.,
\begin{equation}
\ccdf_{\chi^2_{\nu +  2 + 2  k}} \compose \ccdf_{\chi^2_{\nu+2}}^{-1}\left(  p +
d_0  \right)  \,   \geqslant  \,  \ccdf_{\chi^2_{\nu  +  2  +   2  k}}  \compose
\ccdf_{\chi^2_{\nu+2}}^{-1}(p) + d_k.
\label{Eq:Diff2}
\end{equation}
Inserting~\eqref{Eq:Diff2} into~\eqref{Eq:Diff1} allows to conclude.


\section{Construction of analytical multipolar orthonormal basis functions}
\label{App:OrthonormalPolynomials}

Remind that  the search  for an  orthonormal basis  $g_{N,n}$ of  the multipolar
signal space $\E_N$ moves to search for
\begin{equation*}
g_{N,n}(u) = \frac{P_{N,n}(u)}{\left( 1 + u^2\right)^{N + \frac32}},
\end{equation*}
where $\left\{ P_{N,n} \right\}_{n=0}^{2 N}$ is a set of orthonormal polynomials
of degree $n$ for the inner product between polynomials $P, Q$,
\begin{equation*}
\int_\Rset \! P(u) \, Q(u) \, w_N(u) \, du,
\end{equation*}
where the weight $w_N$ of this inner product is given by
\begin{equation*}
w_N(u) =  (1+u^2)^{-2N-3}
\end{equation*}

Under  the lens  of  orthonormal  polynomials, we  then  proceed  in two  steps:
orthogonalization and normalization.


\subsection{Orthogonalization step}
\label{Subsec:OrthogonalisationStep}

The     approach     is     based     on    the     theory     of     orthogonal
polynomials~\cite{Szego1975OrthogonalPolynomials}
(or~\cite[\S~12.1]{Arfhen2013MathPhys}                         or~\cite[Chap.~5,
  \S~2(B)]{Chihara2011OrthogonalPolynomials} for  more recent  references), that
states that, with an inner product with weight $w$ satisfying:
\begin{itemize}
  \item $\left( w p \right)' = w q$ \ with \ $p$ \ polynomials of degree at most
    1, and \ $q$ \ polynomial of  degree at most 2, that rewrites $\dfrac{w'}{w}
    = \dfrac{A}{B}$, with $A  \equiv q - p'$ of degree at most  1, and $B \equiv
    q$ of degree at most 2,
  \item $\displaystyle \lim_{u \to \pm \infty} B(u) \, w(u) = 0$,
\end{itemize}
the functions given  by the so-called Rodrigues' formula  $P_n = \dfrac{c_n}{w}
\dfrac{d^n}{du^n}  \left( B^n  w\right)$ with  $c_n$ normalization  coefficient,
defines a series  of orthogonal polynomials of degree $n$  for the inner product
with weight $w$.  The second condition on the weight  is not always (explicitly)
mentioned in the literature,  but is used in the proof  of the orthogonality via
integration  by  parts.  The  weight  $w_N$  precisely  satisfies  the  previous
conditions with $A = - (4 N + 6) \, u$ polynomials of degree (at most) 1, and $B
= 1 + u^2$, polynomial of degree  (at most) 2, so that, the searched polynomials
are given by
\begin{equation*}
P_{N,n}(u) = c_{N,n} \, (1+u^2)^{2N+3} \, \dfrac{d^n}{du^n} (1+u^2)^{n-2N-3},
\end{equation*}
where $c_{N,n}$ is a normalization coefficient to be found.

Before going further on, it is important to mention that it is generally assumed
that the weight admits moment of  any order, i.e., $\displaystyle \int_\Rset u^n
\, w(u) \, du  < \infty$ for any integer $n$, so that  the series of polynomials
is given for any degree $n$. It is  not the case for $w_N$, which admits moments
only up  to $n  = 4  \, N  + 4$. But  following the  lines of  the proof  of the
Rodrigues' formula, it appears that the approach  is still valid, but only up to
polynomials of degree $n  = 4 \, N + 4$. This is  sufficient here since $n$ will
by limited to $ 2 \, N$.

Let us now introduce  $h: u \mapsto u^{n-2N-3}$ and $g : u  \mapsto 1 + u^2$, so
that  the $n^{\text{th}}$  derivative  term  in the  Rodrigues'  formula is  $(h
\compose      g)^{(n)}$.     We      can     apply      the     the      results
of~\cite{Stanley2024enumerativecombinatorics}   dealing   with   derivative   of
composite function (known as compositional formula, or Fa\`a di Bruno's formula)
\begin{equation*}
(h \mathord\circ g)^{(n)} = \sum_{\pi  \in \Pi_n} h^{(|\pi|)} \mathord\circ g \,
  \prod_{B \in \pi} g^{(|B|)},
\end{equation*}
where  $\Pi_n$ denotes  the set  of partitions  of  $\{ 0,  \ldots ,  n \}$  and
$|\cdot|$  the cardinal  of a  set.  Because  $g^{(k)}  = 0$  for $k  > 2$,  the
non-zero terms are those given by the  partitions with cardinal less or equal to
2, meaning singlets and doublets. Now,  if a partition contains $k$ doublets and
$n-2k$ singlets, then
\begin{equation*}
|\pi|  = n-k.
\end{equation*}
This results in
\begin{equation*}
h^{(|\pi|)}(u) = \dfrac{(-1)^{n-k} \, (2 N + 2 -  k)!}{(2 N + 2 - n)!} \: u^{k -
  2 N - 3}
\end{equation*}
and
\begin{equation*}
\prod_{B \in \pi} g^{(\vert B \vert)}(u) = 2^k \, (2 \, u)^{n - 2 k}.
\end{equation*}
As   there  are   $2k$  variables   to  fix   for  the   doublets  followed   by
$\dfrac{\displaystyle  \prod_{l=0}^{k-1}  \binom{2k  -  2  l}{2}}{k!}$  possible
choices for the doublets, there are
\begin{equation*}
\binom{n}{2 k} \frac{(2 k)!}{2^k k!} \quad \mbox{partitions}
\end{equation*}
to  consider. As  a conclusion,  summing over  all possible  values of  $k$, the
orthonormal polynomial $P_{N,n}$ can be written as:
\begin{equation}
\label{Eq:PNnApp}
P_{N,n}(u) = c_{N,n}
\sum_{k=0}^{\left\lfloor \dfrac{n}{2} \right\rfloor} 
d_{N,n, k} (1+u^2)^k \left( 2u \right)^{n-2k},
\end{equation}
where $\lfloor \cdot \rfloor$ is the floor function and
\begin{equation}
\label{Eq:dNnkApp}
d_{N,n, k} =
\dfrac{(-1)^{n-k} \, n! \, (2N+2-k)!}{(2N+2-n)! \, k! \, (n-2k)!}.
\end{equation}
It follows from Eq.~\eqref{Eq:PNnApp} that the degree of $P_{N,n}$ is indeed $n$.

Note     that      the     searched      $n-$th     derivative      is     given
in~\cite[0.433-3]{Gradshteyn2015tableofintegrals}, where,  in fact,  the maximal
index $k$ in the sum is not expressed.


\subsection{Normalization step}
\label{Subsec:NormalizationStep}

Normalizing the basis means fixing $c_{N,n}$ such that
\begin{equation*}
\int_\Rset P_{N,n}(u)^2 \, w_N(u) \, du = 1.
\end{equation*}

The key idea  here is to introduce  the Rodrigues' formula into  only one factor
$P_{N,n}$ in the integral,
\begin{equation*}
c_{N,n} \int_\Rset P_{N,n}(u) \,  \dfrac{d^n \,  (1+u^2)^{n-2N-3}}{du^n} \, du = 1
\end{equation*}
An integration by parts gives then
\begin{equation*}
1 =  c_{N,n} \left[ P_{N,n}(u) \,  \dfrac{d^{n-1} \, (1+u^2)^{n-2N-3}}{du^{n-1}}
  \right]_{-\infty}^{+\infty}  \!  - c_{N,n}  \!  \int_\Rset  \! P'_{N,n}(u)  \,
\dfrac{d^{n-1} \, (1+u^2)^{n-2N-3}}{du^{n-1}} \, du.
\end{equation*}
Then, the derivative  term can be viewed through the  Rodrigues' formula so that
$\dfrac{1}{(1+u^2)^{2N+2}} \dfrac{d^{n-1}  \, (1+u^2)^{n-2N-3}}{du^{n-1}}$  is a
polynomials  of  degree  $n-1$,  let  say  $Q_{N,n-1}$.   Thus,  $P_{N,n}(u)  \,
\dfrac{d^{n-1}   \,    (1+u^2)^{n-2N-3}}{du^{n-1}}   =    \dfrac{P_{N,n}(u)   \,
  Q_{N,n}(u)}{(1+u^2)^{2N+2}}$.   The numerator  being of  degree  $2 \,  n -  1
\leqslant 4 \,  N -1 < 4 \,  N + 4$, that of the  denominator, the all-inclusive
term cancels out, so that
\begin{equation*}
- \,     c_{N,n}     \int_\Rset     P'_{N,n}(u)     \,     \dfrac{d^{n-1}     \,
  (1+u^2)^{n-2N-3}}{du^{n-1}} \, du = 1.
\end{equation*}
Repeating such an integration by part $n-1$ times more, one achieves
\begin{equation*}
(-1)^n \, c_{N,n} \int_\Rset P^{(n)}_{N,n}(u) \,   (1+u^2)^{n-2N-3} \, du = 1.
\end{equation*}
$P_{N,n}(u)$ being of degree $n$, its $n-$th derivative is constant and writes
\begin{equation*}
P^{(n)}_{N,n}(u) = n! \, p_{N,n},
\end{equation*}
where  $p_{N,n}$ is  the dominant  coefficient of  $P_{N,n}$. The  normalization
coefficient thus satisfies
\begin{equation}
c_{N,n} \, (-1)^n \, n! \,  p_{N,n} \int_\Rset
 (1+u^2)^{n-2N-3} \,  du = 1.
\label{Eq:NormalisationConditionApp}
\end{equation}
The      dominant       coefficient      can      easily       be      extracted
from~\eqref{Eq:PNnApp}-\eqref{Eq:dNnkApp} using binomial formula:
\begin{equation*}
p_{N,n}  =  c_{N,n}  \,  (-1)^n  \,  n!   \sum_{k=0}^{\left\lfloor  \dfrac{n}{2}
  \right\rfloor}  \dfrac{(-1)^k  (2N+2-k)!  \,   2^{n-2k}}{(2N+2-n)!  \,  k!  \,
  (n-2k)!}
\end{equation*}
(noting that $(-1)^{-k} = (-1)^k$). It appears that the sum term is a Gegenbauer
polynomial  $C_n^{(\alpha)} (x)$  evaluated at  $x=1$, for  which an  analytical
formula is available~\cite[Eqs.~22.3.4, 22.4.2]{Abramowitz1972handbook}, leading
to,
\begin{eqnarray}
p_{N,n} & = & c_{N,n} \, (-1)^n \, n! \, C_n^{(2N + 3 - n)}(1)\nonumber\\[2mm]
& = & c_{N,n}\, (-1)^n \, n! \, \binom{4N+5-n}{n}.
\label{Eq:DominantCoefficientApp}
\end{eqnarray}
The remaining integral can be calculated by integration in the complex plane and
from    the   residue    theorem,    the   result    being    given   in    fact
in~\cite[Eq.~8.380-3]{Gradshteyn2015tableofintegrals}, in terms  of $B$ the beta
function,  expression   we  recast   relating  the   Beta  function   $B(a,b)  =
\frac{\Gamma(a)  \Gamma(b)}{\Gamma(a+b)}$  with   the  gamma  function  $\Gamma$
(factorial)~\cite[Eq.~8.384]{Gradshteyn2015tableofintegrals},   the   expression
$\Gamma\left(\frac12\right)                                                    =
\sqrt{\pi}$~\cite[Eq.~8.338-2]{Gradshteyn2015tableofintegrals}  and that  of the
Gamma function of integer argument plus  a half $\Gamma\left( m +\frac12 \right)
=  \frac{\sqrt{\pi}  (2  \,  m)!}{4^m m!}$  (basically  the  so-called  doubling
formula)~\cite[Eq.~8.339-2]{Gradshteyn2015tableofintegrals} to obtain
\begin{eqnarray}
\int_\Rset   (1+u^2)^{n-2N-3}  \,  du   &  =  &  2   \int_0^{+\infty}  
(1+u^2)^{n-2N-3} \, du\nonumber\\[2mm]
& = & B\left( \frac12 \: , \: 2 N + \frac52 - n\right)\nonumber\\[2mm]
& =  & \frac{\Gamma\left(  \frac12 \right) \,  \Gamma\left( 2 N  + \frac52  - n
  \right) }{\Gamma\left( 2 N + 3 - n \right)}\nonumber\\[2mm]
& = & \frac{\pi \, ( 4 \, N + 4 - 2 \, n)! }{4^{2 \, N + 2 - n} \big( (2 \, N +
  2 - n)! \big)^2}.
\label{Eq:ResisualIntegralApp}
\end{eqnarray}
Inserting  Eqs.~\eqref{Eq:DominantCoefficientApp}-\eqref{Eq:ResisualIntegralApp}
into~\eqref{Eq:NormalisationConditionApp} gives the normalization coefficient of
$P_{N,n}$ as:
\begin{equation}
\label{Eq:NormalisationCoefficientApp}
c_{N,n}^2 = \dfrac{4^{2N+2-n} \, (4N+5-2n) \, \big( (2N+2-n)! \big)^2}{\pi \, n!
  \, (4N+5-n)! }.
\end{equation}
As a conclusion, $ \G_N = \Big\{ g_{N,n} \Big\}_{n=0}^{2 N}$ with
\begin{equation}
\label{Eq:SignalBasisON}
g_{N,n}(u) = \dfrac{P_{N,n}(u)}{(1+u^2)^{N+\frac32}}
\end{equation}
and                    $P_{N,n}$                     given                    by
Eqs.~\eqref{Eq:PNnApp}-\eqref{Eq:dNnkApp}-\eqref{Eq:NormalisationCoefficientApp}
is an orthonormal basis of the source space for the natural inner product.


\subsection{Gram-Schmidt equivalence}
\label{Subsec:GSEquivalence}

It is possible  to show that the set of  orthogonal polynomials $\left\{ P_{N,n}
\right\}_{n=0}^{2N}$  obtained previously  coincides exactly  with the  one that
would have  been obtained through  a GS procedure  applied to the  $\F_N$ basis.
Indeed,      it     is      shown     generically      in~\cite[Th.~2.1.1     \&
  \S~2.2]{Szego1975OrthogonalPolynomials} that, given  a weight functions, there
exists a unique orthonormal set of polynomials $P_n,  \: n = 0, \ldots , l$ ($l$
finite or infinite) with respect to the thus defined inner product and such that
$P_n$ is of degree $n$. Since both the GS procedure starting from $x^n, \: n = 0
, \ldots  , l$ thus  ordered and the Rodrigues'  formula gives such  a sequence,
they must coincide.

An  alternative proof  can be  derived  using a  recurrence reasoning:  consider
$\left\{ P_n^{gs} \right\}_{n  = 0}^l$, a set of  orthogonal polynomials, factor
of the weight  $w$ of the inner  product, obtained by a GS  procedure, and $P_n$
that obtained by the Rodrigues' formula.
\begin{itemize}
  \item Initialization step:  $P_0^{gs} = P_0$ because each  one are polynomials
    of degree $0$ and normalized.
  \item Inheritance  stage: Assume $P_i^{gs}  = P_i$  up to a  degree/order $n$.
    $P_{n+1}^{gs}$ is of degree $n+1$ by construction, just like $P_{n+1}$. Each
    of them  has $n+2$ free coefficients.  Being normalized fixes $1$  degree of
    freedom and being orthogonal to $\left\{ P_i^{gs} \right\}_{i=0}^n$ gives $n
    + 1$  additional constraints. As we  have as many constraints  as degrees of
    freedom,  and as  the constraints  are the  same for  both polynomials,  the
    solution can only be unique  thus $P_{n+1}^{gs}$ and $P_{n+1}$ are necessary
    equal.
\end{itemize}


\section{Approximate performance of the receiver based on order selection}
\label{App:PerfCriteria}

Let us restrict the study to the case of  a binary selection $\Mc = M$ vs $\Mc =
M-m$  (typically  with  $M  =  N,  \,   m  =  1$).  Following  the  approach  of
Section~\ref{Subsec:CritreriaOrderSelection}, one can write
\begin{eqnarray*}
\dfrac{\left\|  \widehat{\Ab_M}  \right\|_F^2}{\sigma^2}  & =  &  \dfrac{\left\|
  \xb_M \right\|_F^2}{\sigma^2}\\
& = &  \dfrac{\left\| \xb_{M-m} \right\|_F^2}{\sigma^2} +  \dfrac{\left\| \xb \,
  \gb_{M,m}^\transp \right\|_F^2}{\sigma^2}\\
&  =  &  \dfrac{\left\|  \widehat{\Ab_{M-m}}  \right\|_F^2}{\sigma^2}  +  \Delta
\Crit(M,m) + \delta c(M,m).
\end{eqnarray*}
From  the  gaussianity  and  orthogonality   between  $\xb_{M-m}$  and  $\xb  \,
\gb_{M,m}^\transp$, (conditionally  to $\H_k$) these two  terms are independent:
$\dfrac{\left\|   \widehat{\Ab_{M-m}}   \right\|_F^2}{\sigma^2}$   and   $\Delta
\Crit(M,m) +  \delta c(M,m)$ are  independent and, under $\H_k$,  are noncentral
chi-squared distributed, respectively with $\nu_{M-m}$ and $2 \, d \, m$ degrees
of freedom, and with respective noncentral parameter $k \, \lambda_{M-m}$ and $k
\,  \delta\lambda(M,m)$.  For sake  of  simplicity,  let  us denote  $\X_{M'}  =
\dfrac{\left\|  \widehat{\Ab_{M'}}  \right\|_F^2}{\sigma^2}$,  \ $  \vartheta  =
\dfrac{\eta}{\sigma^2}$  and  $\D  \Crit(M,m)   =  \Delta  \Crit(M,m)  +  \delta
c(M,m)$.  Now, from  $\Mc  = M  \,  \Leftrightarrow \,  \D  \Crit(M,m) >  \delta
c(M,m)$, and omitting the dependence in $M, m$ to lighten the notation, one has
\begin{eqnarray}
\Pr\left[ \left. \left\| \widehat{\Ab_{\Mc}} \right\|_F^2 > \eta \, \right| \,
    \H_k \right] \:
&  = & \: \Pr\left[ \left.  \X_{M-m}  >  \vartheta \,  \right|  \,  \H_k \right]  \,
  \Pr\left[  \left.    \D  \Crit   \leqslant  \delta  c   \,  \right|   \,  \H_k
    \right]\nonumber\\[2mm]
& & \! +  \Pr\left[ \left. \X_{M-m}  + \D  \Crit >
    \vartheta \, , \, \D \Crit > \delta c\, \right| \, \H_k \right],
\label{Eq:ProbaTotales}
\end{eqnarray}
where we  used the independence between  $\X_{M-m}$ and $\D \Crit(M,m)$  for the
first term,  and $\X_M = \X_{M-m}  + \D \Crit(M,m)$  for the second one.  Let us
concentrate on $\H_1$,  i.e., on ${\Pd}_{,\Crit}$; the result  for $\H_0$, i.e.,
for ${\Pfa}_{,\Crit}$, will be similar, just replacing the noncentral parameters
by $0$. Then, the first term writes
\begin{equation}
 \Pr\left[ \left. \X_{M-m}  > \vartheta \, \right| \, \H_1  \right] \, \Pr\left[
   \left.   \D  \Crit   \leqslant  \delta  c  \,  \right|  \,   \H_1  \right]  =
 \cdf_{\chi^2_{2      d       m}\!(\delta      \lambda)}(\delta       c)      \,
 \ccdf_{\chi^2_{\nu_{M-m}}\!\!(\lambda_{M-m})}(\vartheta),
\label{Eq:ProbaTotalesM-m}
\end{equation}
with $\cdf_\zeta  = 1  - \ccdf_\zeta$ denoting  the cumulative  density function
(cdf) of a $\zeta-$distributed random variable. When $\vartheta > \delta c$, the
second one writes
\begin{align}
&\Pr\left[ \left. \X_{M-m} + \D \Crit > \vartheta  \, , \, \D \Crit > \delta c\,
    \right| \, \H_1 \right]\nonumber\\[3mm]
& \qquad  = \Pr\left[ \left.  \D  \Crit > \delta  c\, \right| \, \H_1  \right] -
  \Pr\left[ \left. \X_{M-m}  + \D \Crit \leqslant  \vartheta \, , \,  \D \Crit >
    \delta c\, \right| \, \H_1 \right]\nonumber\\[3mm]
& \qquad = \cdf_{\chi^2_{2 d m}\!(\delta \lambda)}(\delta c) - \int_0^{\vartheta
    - \delta     c}      \!\!\!      \int_{\delta      c}^{\vartheta-u}     \!\!
  \pdf_{\chi^2_{\nu_{M-m}}\!\!(\lambda_{M-m})}(u)    \,     \pdf_{\chi^2_{2    d
      m}\!(\delta\lambda)}(v) \, du \, dv\nonumber\\[3mm]
&  \qquad   =  \:   \ccdf_{\chi^2_{2  d   m}\!(\delta  \lambda)}(\delta   c)  \,
  \ccdf_{\chi^2_{\nu_{M-m}}\!\!(\lambda_{M-m})}(\vartheta-\delta
  c)\nonumber\\[1.5mm]
&  \qquad\quad  +  \int_0^{\vartheta  -   \delta  c}  \!\!   \ccdf_{\chi^2_{2  d
      m}\!(\delta\lambda)}(\vartheta-u)                                       \,
  \pdf_{\chi^2_{\nu_{M-m}}\!\!(\lambda_{M-m})}(u) \, du,
\label{Eq:ProbaTotalesM}
\end{align}
where we use the fact that the inner integral in the second equality is equal to
$\ccdf_{\chi^2_{2   d  m}\!(\delta\lambda)}(\delta   c)  -   \ccdf_{\chi^2_{2  d
    m}\!(\delta\lambda)}(\vartheta-u)$.   The last  result is  still valid  when
$\vartheta     \leqslant     \delta     c$.     In     conclusion,     inserting
Eqs.~\eqref{Eq:ProbaTotalesM-m}-\eqref{Eq:ProbaTotalesM}                    into
Eq.~\eqref{Eq:ProbaTotales} gives
\begin{eqnarray*}
{\Pd}_{,\Crit}(\eta) & = & \cdf_{\chi^2_{2  d m}\!(\delta \lambda)}(\delta c) \,
\ccdf_{\chi^2_{\nu_{M-m}}\!\!(\lambda_{M-m})}(\vartheta)  +  \ccdf_{\chi^2_{2  d
    m}\!(\delta               \lambda)}(\delta               c)               \,
\ccdf_{\chi^2_{\nu_{M-m}}\!\!(\lambda_{M-m})}(\vartheta - \delta c)\\[1.5mm]
&   &    +   \int_0^{\vartheta    -   \delta    c}   \!     \ccdf_{\chi^2_{2   d
    m}\!(\delta\lambda)}(\vartheta-u)                                         \,
\pdf_{\chi^2_{\nu_{M-m}}\!\!(\lambda_{M-m})}(u) \, du.
\end{eqnarray*}
Note that, taking $\delta c = 0$ in Eq.~\eqref{Eq:ProbaTotalesM} one has
\begin{equation*}
\ccdf_{\chi^2_{\nu_M}\!(\lambda_M                 )}(\vartheta)                =
\ccdf_{\chi^2_{\nu_{M-m}}\!\!(\lambda_{M-m}  )}(\vartheta)   +  \int_0^\vartheta
\ccdf_{\chi^2_{2         d          m}\!(\delta\lambda)}(\vartheta-u)         \,
\pdf_{\chi^2_{\nu_{M-m}}\!\!(\lambda_{M-m} )}(u) \, du,
\end{equation*}
which, written at $\vartheta - \delta c$,  finally gives
\begin{eqnarray}
{\Pd}_{,\Crit}(\eta) & = & \cdf_{\chi^2_{2  d m}\!(\delta \lambda)}(\delta c) \,
\ccdf_{\chi^2_{\nu_{M-m}}\!\!(\lambda_{M-m})}(\vartheta)      \:       +      \:
\ccdf_{\chi^2_{2      d       m}\!(\delta      \lambda)}(\delta       c)      \,
\ccdf_{\chi^2_{\nu_M}\!\!(\lambda_M)}(\vartheta - \delta c)\nonumber\\[1.5mm]
&   &  +   \int_0^{\vartheta  -   \delta   c}  \!    \Big(  \ccdf_{\chi^2_{2   d
    m}\!(\delta\lambda)}(\vartheta-u) \nonumber\\[1.5mm]
& & \qquad - \ccdf_{\chi^2_{2 d m}\!(\delta \lambda)}(\delta c) \ccdf_{\chi^2_{2
    d   m}\!(\delta\lambda)}(\vartheta    -   \delta    c   -   u)    \Big)   \,
\pdf_{\chi^2_{\nu_{M-m}}\!\!(\lambda_{M-m})}(u) \, du,
\label{Eq:PdC}
\end{eqnarray}
and similarly, with central chi-squared  distributions instead of the noncentral
ones for ${\Pfa}_{,\Crit}(\eta)$.

From this expression, when $\eta = \sigma^2 \vartheta$ is fixed,
\begin{itemize}
\item  When $\delta  c \to  +\infty$, we  have $\left(  {\Pfa}_{,\Crit} \,  , \,
  {\Pd}_{,\Crit} \right) \to \left( {\Pfa}_{,M-m} \, , \, {\Pd}_{,M-m} \right)$,
  which  is in  accordance  with the  fact  that when  the  penalization in  the
  criterion is too large, the lowest  order is ``almost always'' chosen. This is
  roughly the case for $\Crit = \BIC$.
\item  When  $\delta  c  \to  0$,  we  have  $\left(  {\Pfa}_{,\Crit}  \,  ,  \,
  {\Pd}_{,\Crit} \right)  \to \left(  {\Pfa}_{,M} \,  , \,  {\Pd}_{,M} \right)$,
  which is  the opposite situation. With  a weak penalization in  the criterion,
  since the  main term (the projection  energy) increases with $M$,  the highest
  order is ``almost always'' chosen.
\end{itemize}
$\Crit = \AIC$ is a situation between both extreme cases.

Now, for fixed $\delta c$, it appears that
when $\vartheta <  \delta c$, one has from Eq.~\eqref{Eq:PdC},  with central law
instead  of noncentral  ones,  that ${\Pfa}_{,\Crit}(\eta)  = \cdf_{\chi^2_{2  d
    m}}(\delta c) \,  \ccdf_{\chi^2_{\nu_{M-m}}}(\vartheta) + \ccdf_{\chi^2_{2 d
    m}}(\delta c)$, that gives, fixing the false alarm probability,
\begin{equation*}
\vartheta    =    \ccdf_{\chi^2_{\nu_{M-m}}}^{-1}     \left(    \dfrac{\Pfa    -
  \ccdf_{\chi^2_{2 d m}}(\delta c)}{\cdf_{\chi^2_{2 d m}}(\delta c)} \right).
\end{equation*}
Now,  from $\dfrac{\Pfa  - \ccdf_{\chi^2_{2  d m}}(\delta  c)}{\cdf_{\chi^2_{2 d
      m}}(\delta   c)}   =   \Pfa   -   \dfrac{\ccdf_{\chi^2_{2   d   m}}(\delta
  c)}{\cdf_{\chi^2_{2  d  m}}(\delta  c)}  \left(  1 -  \Pfa  \right)$,  $  1  =
\ccdf_{\chi^2_{\nu_M}\!\!(\lambda_M)}  \compose \ccdf_{\chi^2_{\nu_M}}^{-1}(\Pfa
+ 1 - \Pfa)$ we have, from an order 1 Taylor expansion around $\Pfa$,
\begin{equation*}
{\Pd}_{,\Crit}(\Pfa) \,  = \, \cdf_{\chi^2_{2 d  m}\!(\delta \lambda)}(\delta c)
\, {\Pd}_{,M-m}(\Pfa) \, + \, \ccdf_{\chi^2_{2 d m}\!(\delta \lambda)}(\delta c)
\, {\Pd}_{,M}(\Pfa) \, + \, O\left( 1 - \Pfa \right).
\end{equation*}
The  leading term  is nothing  but a  convex combination  of ${\Pd}_{,M-m}$  and
${\Pd}_{,M}$, which allows  to conclude that for  ``sufficiently large'' $\Pfa$,
the criterion leads to a trade off  between the performance of the $M-$order and
$(M-m)-$order receivers.  The  weight of the convex combination  depends both on
the energy proportion  between $M$ and $M-m$ (see non-central  parameter) and on
the criterion (see argument). Typically, for $m =  1$ (and $M = N = 2$ and $\SNR
= -22$~dB),  for AIC  one has  $\cdf_{\chi^2_{2 d}\!(\delta  \lambda)}(\delta c)
\approx .642$ for $\sce{1}$ and $\cdf_{\chi^2_{2 d}\!(\delta \lambda)}(\delta c)
\approx  .8817$  for  $\sce{2}$,  showing   that  in  the  second  scenario  the
performance is closer  to that of the  dipolar one than for  the first scenario.
For BIC,  $\cdf_{\chi^2_{2 d}\!(\delta \lambda)}(\delta  c) \approx 1$  for both
scenarios,  in accordance  to the  ``almost systematic''  choice of  the dipolar
receiver.


\section{Dealing with colored noise}
\label{App:ColoredNoise}

Whereas  we  made the  very  strong  assumption that  the  noise  was white  and
Gaussian,      in     operational      conditions,     the      whiteness     is
contestable~\cite{Ash1997Noise,  Sheinker2007Colorednoise}.  However,  the  main
results we derived remain valid.

In the case  where the spatial and temporal correlation  structures of the noise
are  decoupled, which  is denoted  $  \nb \sim  \N_{d,K}\left( \boldsymbol{0}  ,
\Sigmab_s \otimes \Sigmab_t \right)$, where  the $d \times d$ symmetric positive
definite matrix  $\Sigmab_s$ represents  the spatial covariance  (structure) and
the $K  \times K$  symmetric positive definite  matrix $\Sigmab_t$  the temporal
ones,    in   expression~\eqref{Eq:GaussianMatrixPDF}    $\frac1{\sigma^2}   \yb
\yb^\transp$   is   replaced   by    $\Sigmab_s^{-1}   \yb   \,   \Sigmab_t^{-1}
\yb^\transp$~\cite[Def.~2.2.1 \&  Th.~2.2.1]{Gupta2018matrix}. Writing  the GLRT
is in  fact equivalent to  first perform a noise  whitening, i.e., to  work with
$\Wb_{\!\!  s}  \, \xb  \, \Wb_{\!\!   t}$ where $\Wb_{\!\!  s} \,  \Sigmab_s \,
\Wb_{\!\!   s}^\transp =  \Ib_d$ and  $\Wb_{\!\!  t} \,  \Sigmab_t \,  \Wb_{\!\!
  t}^\transp = \Ib_K$  (e.g., obtained via a diagonalization, or  via a Cholesky
decomposition of the covariance matrices~\cite{Golub2013matrixcomputation}). If,
indeed, from~\cite[Th.~2.3.10]{Gupta2018matrix} the  noise part $\widetilde{\nb}
=     \Wb_{\!\!     s}     \,    \nb     \,    \Wb_{\!\!      t}$    has     the
distribution~\eqref{Eq:GaussianMatrixPDF},    the    signal    part    is    now
$\widetilde{\sb} = \Wb_{\!\! s}  \, \sb \, \Wb_{\!\! t} =  \Wb_{\!\! s} \, \Ab_N
\Fb_N \,  \Wb_{\!\! t}$.  The  signal obtained  after whitening the  noise still
decomposes on a basis, now $\widetilde{\Fb}_N  = \Fb_N \Wb_{\!\!  t}$, where the
coefficients of the decomposition are given by $\widetilde{\Ab}_N = \Wb_{\!\! s}
\, \Ab_N$. All the analyses of the paper remain thus valid, except that we loose
the analytical construction of the orthonormal basis. Indeed, a short inspection
allows  to  see that  $\Gb_N  \,  \Wb_{\!\!  t}$  is  not  Stiefel, except  when
$\Wb_{\!\! t}  \propto \Ib$ (the  noise is  temporally white). However,  one can
have  either recourse  to a  Gram-Schmidt orthonormalization  procedure (or  any
other  orthonormalization  one),  with  the the  consequence  of  an  additional
computational cost and possible  numerical instability issues. Alternatively, we
can    still     implement    an     $M-$order    receiver    of     the    form
Eqs~\eqref{Eq:AMLE}-\eqref{Eq:Energy_test},  with basis  $\widetilde{\Fb}_M$ and
its Moore-Penrose pseudo-inverse.

Finally, when the  spatial and temporal correlation are  coupled, one vectorizes
the observation $\vec  \left( \xb^\transp \right) =  \begin{bmatrix} \xb_{1,1} &
  \cdots & \xb_{1,K} & \cdots & \xb_{d,K} \end{bmatrix}^\transp $, this having a
$ d  K \times d K$  non-Kronecker product-form covariance matrix  $\Sigmab$. One
can still  whitening the noise by  working with $ \Wb  \! \vec\left( \xb^\transp
\right)$ where $\Wb  \, \Sigmab \, \Wb^\transp = \Ib_{d  K}$.  Since $\vec\left(
\Fb_N^\transp  \Ab_N^\transp  \right)  =  \left(  \Ib_d  \otimes  \Fb_N  \right)
\vec\left( \Ab_N^\transp  \right)$~\cite[Chap.~2, Sec.~4]{Magnus2019matrix}, one
can      implement     an      $M-$order      receiver      of     the      form
Eqs~\eqref{Eq:AMLE}-\eqref{Eq:Energy_test}  where  $\Ab_M$  is replaced  by  the
vectorization  $\vec\left(  \Ab_M^\transp  \right)$   and  with  basis  $\Wb  \!
\left(\Ib_d \otimes \Fb_M^\transp \right)$ and its Moore-Penrose pseudo-inverse,
or  a Gram-Schmidt  procedure  can be  applied to  $\Wb  \! \left(\Ib_d  \otimes
\Fb_M^\transp \right)$.

\end{appendices}



\section*{Declarations}

\bmhead{Author contribution} CCP  \& SZ contributed to the original  idea and to
the proofs  and mathematical developments.   LLR, OP  and RK contributed  to the
magnetism and MAD principles and OJJM contributed to the signal processing ones.
All  the authors  contributed to  the discussion  behind this  work, and  to the
writing and review of the manuscript.

\bmhead{Funding} This work was partially supported by Naval Group as part of the
joint laboratory on naval electromagnetism grouping the involved partner of this
paper.

\bmhead{Ethics approval and consent to participate} Not applicable.

\bmhead{Consent for publication} Granted.

\bmhead{Data material and code availability} Not applicable.

\bmhead{Competing interests} The  authors declare that they have  no conflict of
interest.


\bibliography{references}

\end{document}